\documentclass{article}

\usepackage{arxiv}

\usepackage[utf8]{inputenc} 
\usepackage[T1]{fontenc}    
\usepackage{hyperref}       
\usepackage{url}            
\usepackage{nicefrac}       
\usepackage{microtype}      
\usepackage{subcaption}
\usepackage{mathrsfs}
\usepackage{times}
\usepackage{amsmath,amssymb,amsfonts}
\usepackage{epsfig}
\usepackage{graphicx}
\usepackage{subfloat}
\usepackage{hhline}
\usepackage{color}
\usepackage{multirow}
\usepackage{threeparttable}
\usepackage{booktabs}
\usepackage{diagbox}
\usepackage{algorithmic}
\usepackage{bm}
\usepackage[ruled,linesnumbered]{algorithm2e}
\usepackage{caption}
\usepackage[dvipsnames]{xcolor}  
\usepackage{listings}  
\usepackage{bbding}
\newcommand*{\affmark}[1][*]{\textsuperscript{$#1$}}

\title{Mini Honor of Kings: A Lightweight Environment for Multi-Agent Reinforcement Learning}

\author{
Lin Liu$^1$, Jian Zhao$^{2}$, Cheng Hu$^{4}$, Zhengtao Cao$^{2}$, Youpeng Zhao$^{3}$, Zhenbin Ye$^{1}$, \\
\textbf{Meng Meng$^{1}$, Wenjun Wang$^{1}$, Zhaofeng He$^{4}$, Houqiang Li$^{3}$, Xia Lin$^{1}$, Lanxiao Huang$^{1}$}\\
\affmark[1]Tencent Timi Studio \\
\affmark[2]Polixir\\
\affmark[3]University of Science and Technology of China\\
\affmark[4]Beijing University of Posts and Telecommunications\\
\texttt{\{lincliu,zhenbinye,promengmeng\}@tencent.com} \\
\texttt{\{jamesonwang,hugolin,jackiehuang\}@tencent.com} \\
\texttt{\{jian.zhao,zhengtao.cao\}@polixir.ai}\\
\texttt{zyp123@mail.ustc.edu.cn}, \texttt{lihq@ustc.edu.cn}\\
\texttt{hucheng@bupt.edu.cn}, \texttt{zhaofenghe@bupt.edu.cn}\\
}

\begin{document}
\maketitle
\begin{abstract}
Games are widely used as research environments for multi-agent reinforcement learning (MARL), but they pose three significant challenges: limited customization, high computational demands, and oversimplification.
To address these issues, we introduce the first publicly available map editor for the popular mobile game Honor of Kings and design a lightweight environment, Mini Honor of Kings (Mini HoK), for researchers to conduct experiments. 
Mini HoK is highly efficient, allowing experiments to be run on personal PCs or laptops while still presenting sufficient challenges for existing MARL algorithms.
We have tested our environment on common MARL algorithms and demonstrated that these algorithms have yet to find optimal solutions within this environment.
This facilitates the dissemination and advancement of MARL methods within the research community.
Additionally, we hope that more researchers will leverage the Honor of Kings map editor to develop innovative and scientifically valuable new maps.
Our code and user manual are available at: https://github.com/tencent-ailab/mini-hok.

\end{abstract}


\section{Introduction}
Cooperative multi-agent systems are essential in numerous real-world scenarios where a group of agents collaborate to accomplish a task and maximize cumulative rewards for the team. \cite{busoniu2008comprehensive, tuyls2012multiagent}.
The emergence of reinforcement learning techniques \cite{silver2016mastering, vinyals2019grandmaster, berner2019dota} has propelled the progress of multi-agent reinforcement learning (MARL) across an array of domains, including autonomous vehicles \cite{cao2012overview}, traffic light control \cite{wu2020multi}, robotics \cite{al2019deeppool} and the smart grid \cite{zhang2021decentralized}.
However, for algorithm research, games are still the most commonly utilized experimental environment for MARL.

As an experimental environment, the game environment offers several inherent advantages. 
Firstly, games are inherently engaging and can easily capture the interest of users. 
Secondly, popular games typically have large user bases and have undergone extensive testing, resulting in robust environments that are less prone to bugs.
Lastly, being designed for general consumers, games are intuitive and do not require extensive domain knowledge to understand.
The MARL research encompasses a variety of game environments, including Hanabi Learning Environment \cite{bard2020hanabi}, PettingZoo \cite{terry2021pettingzoo}, StarCraft Multi-Agent Challenge (SMAC) \cite{samvelyan2019starcraft} and Google Research Football (GRF) \cite{kurach2020google}.

As the number of MARL environments grows, the evaluation system for these environments must be correspondingly enhanced and refined.
We propose that a purely research-oriented game environment should embody the following characteristics:
\begin{itemize}
\item Lightweight: Given that reinforcement learning necessitates the collection of vast amounts of data, the interaction between the agent and the environment must be efficient. 
\item Potential: While maintaining a lightweight design, the environment should not be overly simplistic. 
An environment that allows existing MARL algorithms to quickly converge to a global optimal solution fails to effectively evaluate these algorithms.
\item Customizable: Different MARL algorithms often address various challenges within multi-agent systems. 
Therefore, the environment should include an editor that allows for the modification and customization of its elements.
\end{itemize}
Unfortunately, existing game environments often struggle to simultaneously meet all these criteria.

In this paper, we introduce the latest Honor of Kings map editor.
Leveraging this powerful tool, researchers can design a variety of engaging tasks, such as quick search, tower defense and boss battle.
These generated environments support fast execution without the need for rendering an interface on Linux or Windows operating system. 
Upon completion, the runs produce video files that can be played back using specific software for visual review.
A notable advantage of this map editor over others is independence on a UI interface for modifications, which can be inconvenient and restrict flexibility in parameter adjustments. 
Instead, users can simply adjust the configuration files to change the number, type, and properties of agents in the environment.
This feature facilitates convenient training and learning for tasks such as curriculum learning and transfer learning.

\begin{figure}[htbp]
	\centering
	\subfloat[]{
		\includegraphics[width=0.6\textwidth]{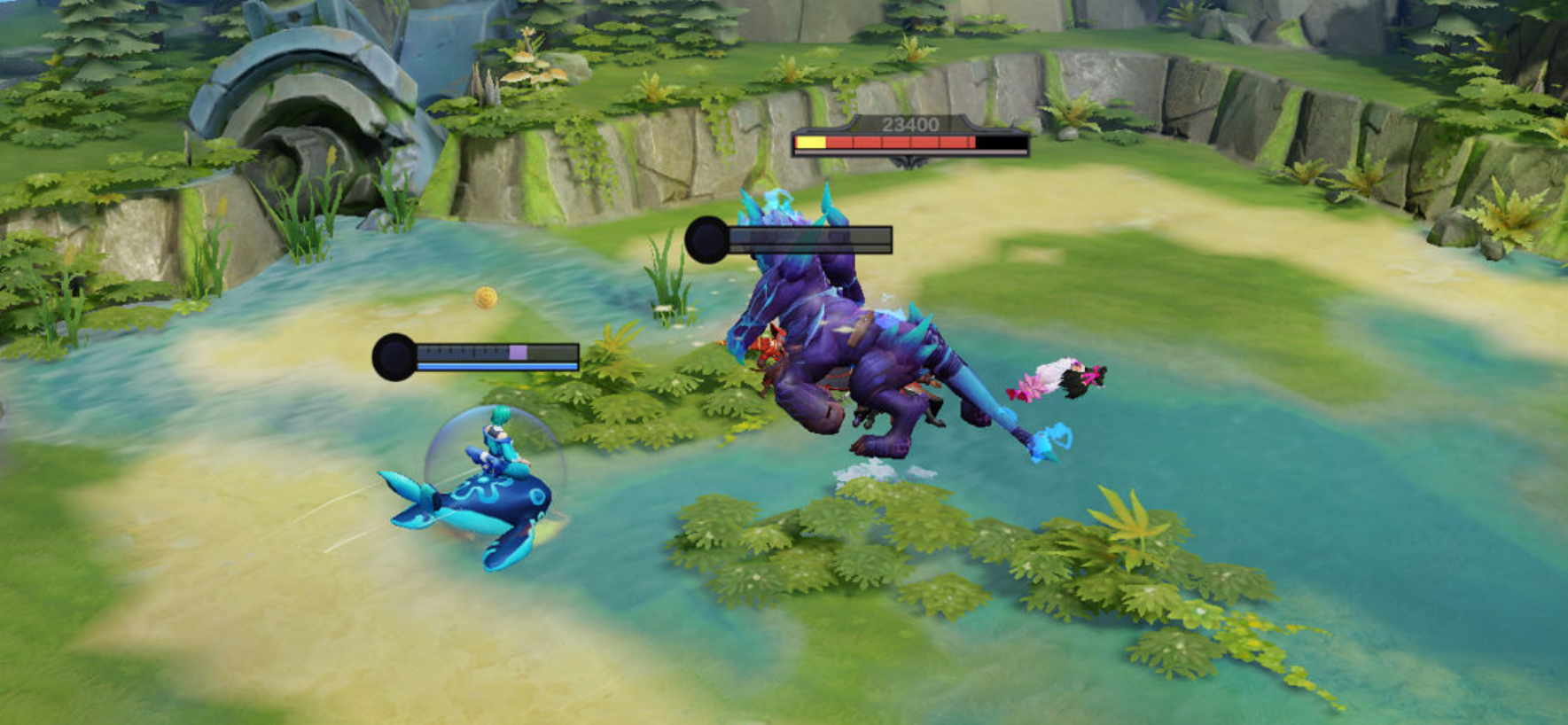}
		\label{example}
	}
	\subfloat[]{
		\includegraphics[width=0.32\textwidth]{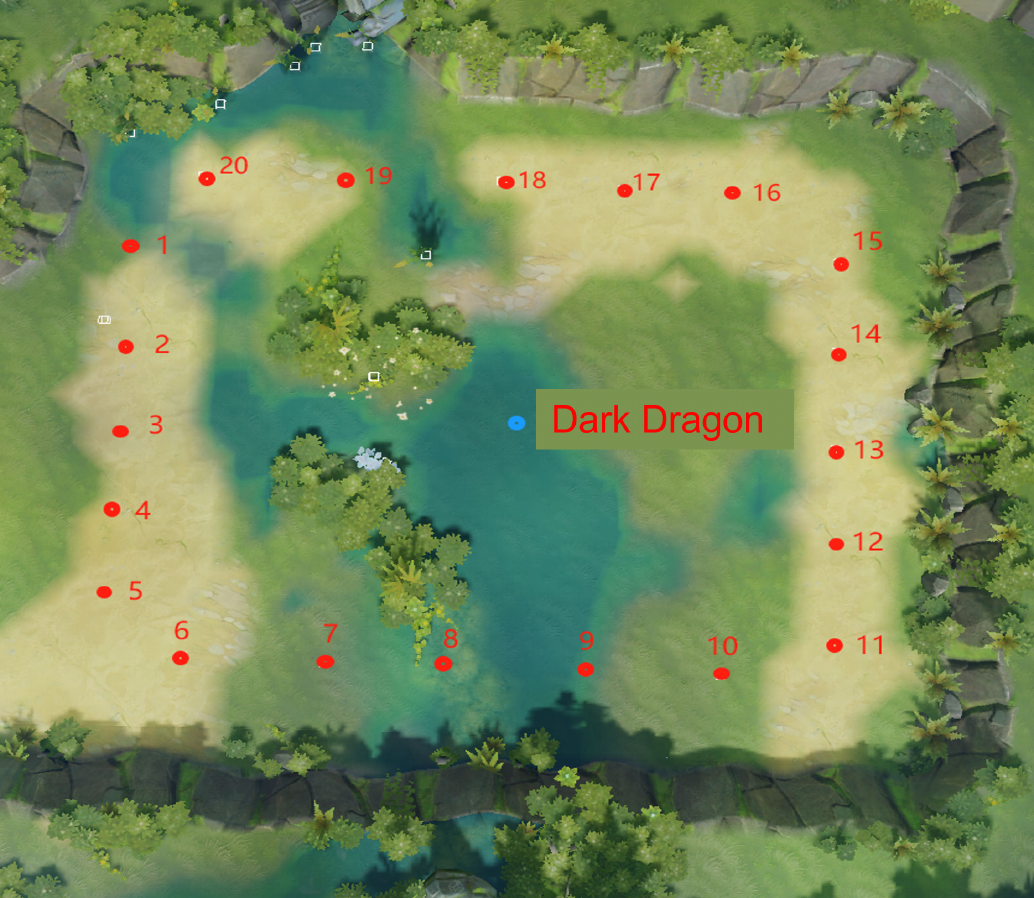}
		\label{init_loc}
	}
	\caption{Examples to show characteristics of our environment. (a) An example to show the battling scenario of our Mini HoK environment, where heroes are cooperating to fight against the Dark Dragon. (b) An example to present the start positions of heroes in the environment. The default start positions are designated as points 1-5. The middle point is initial position of the Dark Dragon.
	}
\end{figure}

With this tool, we have designed a new MARL environment, Mini Honor of Kings (Mini HoK), based on the popular game Honor of Kings.
Prior to this, Honor of Kings had introduced two environments: Hok Arena (1v1) \cite{wei2022honor} and Hok 3v3 \cite{liu2023hokv}. 
However, these environments are aimed at researchers with a solid foundation in MARL and the computational resources for in-depth scientific research.
Mini HoK is set in the classic ``Fighting the Dark Dragon" scenario from Honor of Kings. 
Users can choose any number and type of heroes to participate, with the objective being to cause as much damage as possible on the Dark Dragon within a specified time frame. 
In this paper, we have designed 7 different modes, each corresponding to a specific setting, to facilitate users in testing the performance of their algorithms.

We have evaluated common MARL algorithms in this environment and observed efficient interaction between the environment and the agents. 
Using a Nvidia GeForce RTX 4090 GPU and a 4-core CPU for training, one episode can be completed every second. 
This efficiency makes the environment highly suitable for educational purposes and for beginners.
Moreover, this environment has demonstrated significant potential. 
The performance of common MARL algorithms in the Mini HoK environment still lags behind that of a rule-based method. 
This gap presents an opportunity for researchers to explore and develop more effective MARL algorithms within this environment.

In addition to supporting the study of general MARL algorithms, this environment simplifies research on curriculum reinforcement learning and transfer reinforcement learning.
We have open-sourced the entire environment and provided user manual at https://github.com/tencent-ailab/mini-hok.
Looking ahead, we hope contributions from the community to enhance various algorithms within the Mini Honor of Kings environment. 
We also encourage the use of the map editor to design more interesting and scientifically meaningful game environments.


\section{Related Work}
A variety of reinforcement learning environments have played a pivotal role in advancing the study of multi-agent reinforcement learning (MARL).
However, existing benchmarks own some limitations.
To address these drawbacks, we propose the novel Mini Honor of Kings Environment for further MARL research.

Firstly, with the development of MARL algorithms demonstrating increasingly superior performance, quite a few benchmark environments have become easy to solve.
For instance, the cooperative scenarios provided by OpenSpiel \cite{lanctot2019openspiel} and PettingZoo \cite{terry2021pettingzoo} are set in grid-worlds and board games, which are characterized by simplistic dynamics or a limited number of agents, rendering them susceptible to facile solutions.
Moreover, in the mainstream MARL benchmark, SMAC \cite{samvelyan2019starcraft}, recent work has reported remarkable achievements, even near-perfect win rates, across most scenarios \cite{wang2020qplex, yu2022surprising, kuba2021trust}.
In light of these developments, there arises a pressing need for the creation of a novel MARL benchmark environment capable of presenting substantial challenges, thereby facilitating the discernment of algorithmic performance.

What's more, many prevalent MARL benchmark environments are based on video games or intricate simulators.
For example, SMAC is built on the game Starcraft II, GRF is rooted in  the GameplayFootball simulator\cite{kurach2020google} and Multi-Agent Mujoco \cite{de2020deep} represents a multi-agent adaptation of the MuJoCo environment \cite{todorov2012mujoco} tailored for robotic control.
These benchmarks are characterized by the complex engine featuring multiple agents, intricate rules, and dynamic interactions.
Consequently, training MARL algorithms within these environments are computationally intensive and time-consuming, thereby curtailing accessibility for researchers with limited resources.
Furthermore, this inherent complexity poses challenges in scaling up the number of agents due to computational constraints.
Conversely, our Mini Honor of Kings Environment possesses exceptional computational efficiency, enabling researchers to conduct experiments even with limited computational resources.

Last but not least, certain advanced simulators impose restrictive use terms or necessitate access to closed-source binaries, thereby limiting the ability to modify the environments \cite{todorov2012mujoco, vinyals2017starcraft}.
What's worse, some environments have witnessed maintenance cessation, exemplified by the discontinuation of updates for SMAC due to the cessation of updates for the StarCraft II game. 
These factors collectively poses challenges to editing the environments for researchers, rendering the scenarios outdated and unable to cater to the evolving and diverse requirements of future research.
Conversely, our environment is built on the game Honor of Kings, known as one of the world's most popular and high-grossing games of all time.
The continuous updates of the original game ensure ongoing support, enabling developers to facilitate the development of this benchmark. 
Moreover, the endeavor of continuous benchmark environment updating incurs significant costs, underscoring the need for a benchmark environment that supports user customization.
In this regard, our work offers a tool that enables users to modify environments, thus facilitating the generation of diverse scenarios aligned with research requirements.

Apart from our proposed environment, another reinforcement learning benchmark environment, named Honor of Kings Arena \cite{wei2022honor}, is also derived from the game Honor of Kings. 
This environment closely resembles the original game dynamics, where players control heroes to gain gold and experience by eliminating other in-game units.
The primary objective is to destroy opponent turrets and the base crystal while safeguarding one's own structures.
However, Honor of Kings Arena is a 1v1 environment, predominantly catering to single-agent reinforcement learning rather than cooperative multi-agent systems.
Although the developers have recently updated the Hok3v3 environment, its complexity is heightened due to the diverse array of units and intricate map configurations.
Consequently, the computational demands for training are substantial, potentially exceeding the resources available to common researchers.
In contrast, our Mini Honor of Kings Environment is centered on the task of combating the Dark Dragon, offering a streamlined and efficient training process conducive to convenient experiments.

\section{Honor of Kings Map Editor}
As the foundation of our Mini HoK environment, the Honor of Kings Map Editor allows users to freely customize scenarios.
As depicted in Figure~\ref{init_loc}, there are twenty initial positions where heroes can be placed, providing users with the flexibility to determine start positions using the map editor. 
Moreover, users can configure the behavior of agents, thereby altering the objectives of the scenario. 
For example, setting the dragon to stay still can transform the scenario into a navigation task, where agents have to reach targeted positions. 
Although we provide tasks where agents cooperate to fight the dragon, the editor enables users to create a wide range of tasks for MARL research.

\begin{figure*}[htbp]
	\centering
	\includegraphics[width=0.95\textwidth]{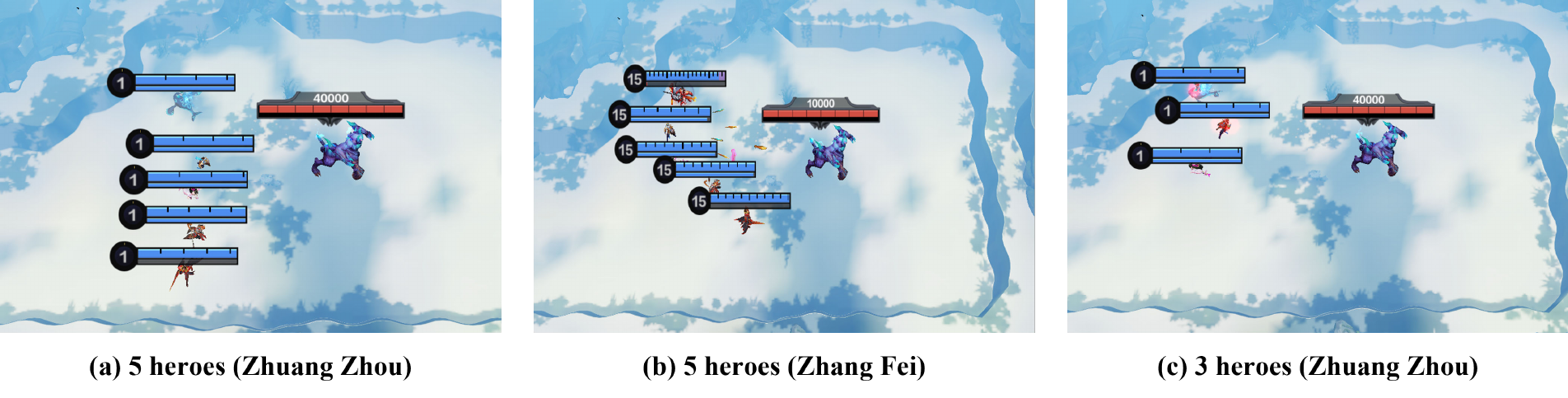}
	\caption{Examples to show the expansibility of our Mini Honor of Kings Environment. In figure (a) and (b), five heroes engage in battle against dragon, featuring Zhuang Zhou in the former and Zhang Fei in the latter. Figure (c) showcases a scenario with three heroes participating in combat.  Also, the levels of the heroes and the dragon's health points can be adjusted. Additionally, users can toggle the presence of equipment for the heroes or other parameters within our environment, thereby offering a comprehensive array of customization options.
    }
	\label{example2}
\end{figure*}

The map editor endows our environment with high expansibility, allowing users to modify various aspects of the settings.
Figure~\ref{example2} demonstrates this trait clearly, showcasing the ability to adjust parameters such as the number and identity of heroes, their respective levels, and the health points of the dragon. 
Furthermore, although not depicted in the figure, users retain the option to determine whether heroes are equipped or other parameters.
In comparison to existing benchmarks like SMAC where scenario editing is feasible, our map editor offers more convenience and autonomy, as SMAC only allows users to manipulate the number and types of units rather than changing specific attributes.

\section{Mini Honor of Kings Environment} \label{environment}

\begin{table}[htbp]
\centering
\renewcommand{\arraystretch}{1.3}{
    \begin{tabular}{|c|c|c|c|c|c|c|}
    \hline
Setting     &   Hero level & Skill 1 & Skill 2  & Skill 3 & Equipment &  \multicolumn{1}{l|}{Hero composition} \\ \hline
(A)   & 1 & 0   & 0  & 0 & \XSolid     & basic    \\  \hline
(B)   & 4 & 0   & 0  & 0 &  \XSolid    & basic    \\ \hline
(C)   & 15 & 0  & 0  & 0 &  \XSolid    & basic    \\ \hline
(D)   & 15 & 6  & 6  & 3 &  \XSolid    & basic    \\ \hline
(E)   & 15 & 6  & 6  & 3 &   \XSolid   & homo (Sun Wukong) \\ \hline
(F)   & 4 & 2   & 1  & 1 &  \XSolid    & basic    \\ \hline
(G)   & 4 & 2   & 1  & 1 &  \Checkmark  & basic      \\ \hline
\end{tabular}}
    \caption{Different configurations of the experiments. Table entries denote the levels of heroes and their respective three skills. 
    The symbol \XSolid~indicates absence of equipment for all heroes, while \Checkmark~signifies all heroes possess six pre-set equipment, as determined by the game engine. 
    When all skill levels are set to zero, it implies that the heroes are only capable of executing normal attacks.
    The basic hero configuration includes the five aforementioned heroes, while the homogeneous hero composition consists of five Sun Wukong for simplicity.}
    \label{tab1}
\end{table}

Our Mini Honor of Kings Environment is based on the engine offered by Honor of Kings and accessible to all individuals for non-commercial activities.
This environment consists of two components to interact with the gamecore, namely client and server.
The server hosts the map editor and game engine, allowing users to input starting command or change environment configurations when simulating battles between heroes and the Dark Dragon.
On the other hand, users can develop their programs within the client.
To aid users in this process, we provide example code based on the Pymarl2 algorithm library \cite{hu2021rethinking}, demonstrating the application of above mentioned classic MARL algorithms within our benchmark environment.

\subsection{Environment Settings}

\noindent\textbf{Overview} The simulated environment portrays a scenario wherein heroes, under the control of a multi-agent system, collaborate to combat the Dark Dragon. 
Controlled by heuristic rules characterized by significant stochasticity, the Dark Dragon possesses three skills whose release occurs randomly. 
By default, the combat involves five heroes, each representing different hero types and possessing distinct attributes. 
The basic attributes of each hero contains health point volume, magical point volume, attack, defence, resistance and so on.
Additionally, the game's difficulty level can be modified by adjusting parameters such as the health points of the dragon or specifying whether the heroes have equipment.


\noindent\textbf{State \& Observations} We delineate the concept of state as the comprehensive set of data provided by the environment subsequent to action execution.
Meanwhile, observations, defined as inputs for the system, consist of individual observations and joint observations, where the former is the input for each agent while the latter involves the amalgamation of individual observations.
Notably, the joint observation feature is only accessible to agents during centralised training for value estimation.
Leveraging the state information, users can customize and redefine their observations according to their specific needs.
In this work, we furnish rudimentary observation features for each agent, encompassing the agent's own location and health point volume, as well as those of the Dark Dragon. 
This simplistic design aims to accommodate diverse settings irrespective of the hero selections and showcase the extensibility of our environment.

\noindent\textbf{Actions} The available actions for an individual agent encompass movement in eight directions and various attack maneuvers, including standard attack, three hero-specific skills and a summoner skill, which is selected according to specific rules for each hero.
Hero skills can be categorized based on their release mechanisms, including targeting specific entities, directions, or positions. 
To streamline the action space of agents, users can modify functions that transform actions into commands required by the gamecore, implementing rules to account for the unique skill sets of each hero.
Given the inherent complexity arising from the distinct skill sets possessed by each hero, such adjustments are helpful for managing the intricate nature of the problem. 
In our current configuration, skills are by default directed towards the dragon for simplicity. 
To enhance the complexity of the problem, considerations such as the existence of additional adversaries or supportive skills for teammates can be integrated into the action space, thereby expanding the range of possible actions.
Furthermore, both hero skills and summoner skills have cooldown time, which may render certain actions temporarily unavailable. 
Utilizing the state information provided by the environment, legal actions features can be derived to judge which actions are permissible for selection.

\noindent\textbf{Rewards} 
In our setup, we provide a basic configuration for dense reward.
To be specific, the reward computation at each time step relies on the difference in the health points of the dragon, adjusted by a factor of -0.01 to reflect the decrement in current health points.
This configuration gives dense feedback to the system, facilitating the models in discerning actions conducive to inflicting damage upon the dragon. 
Moreover, users can use the reduction in the agents' health points as a signal to encourage the agents to evade enemy attacks.
In situations that require sparse rewards, the outcome of defeating the dragon can serve as a viable reward signal.
Leveraging the state information, rewards can be tailored by users to accommodate a spectrum of research requirements.


\noindent\textbf{Stochasticity} 
The stochastic nature of the environment in reinforcement learning plays a pivotal role by introducing variability and uncertainty into the interactions of agents, thereby incentivizing exploration and aiding in the generalization and adaptation of agents.
Our environment possesses enough stochasticity, providing an ideal platform for research endeavors.
For example, being built on the game Honor of Kings, the heroes' skills may have some random characteristics and critical strikes may occur. 
Similarly, the Dark Dragon's skill release also exhibit randomness, further contributing to the unpredictability of the environment.
Additionally, users retain the option to generate randomized starting positions tailored to their specific experimental needs.


\subsection{APIs and Implementation}

Composed of both client and server components, the environment transmits data through JSON strings for communication purposes.
To enhance usability, we encapsulate the interaction with the environment using Python.
An illustrative example of the interface and config is provided in Figure~\ref{config}, accompanied by descriptions of key functions outlined below:

\begin{figure*}[htbp]
	\centering
	\includegraphics[width=0.8\textwidth]{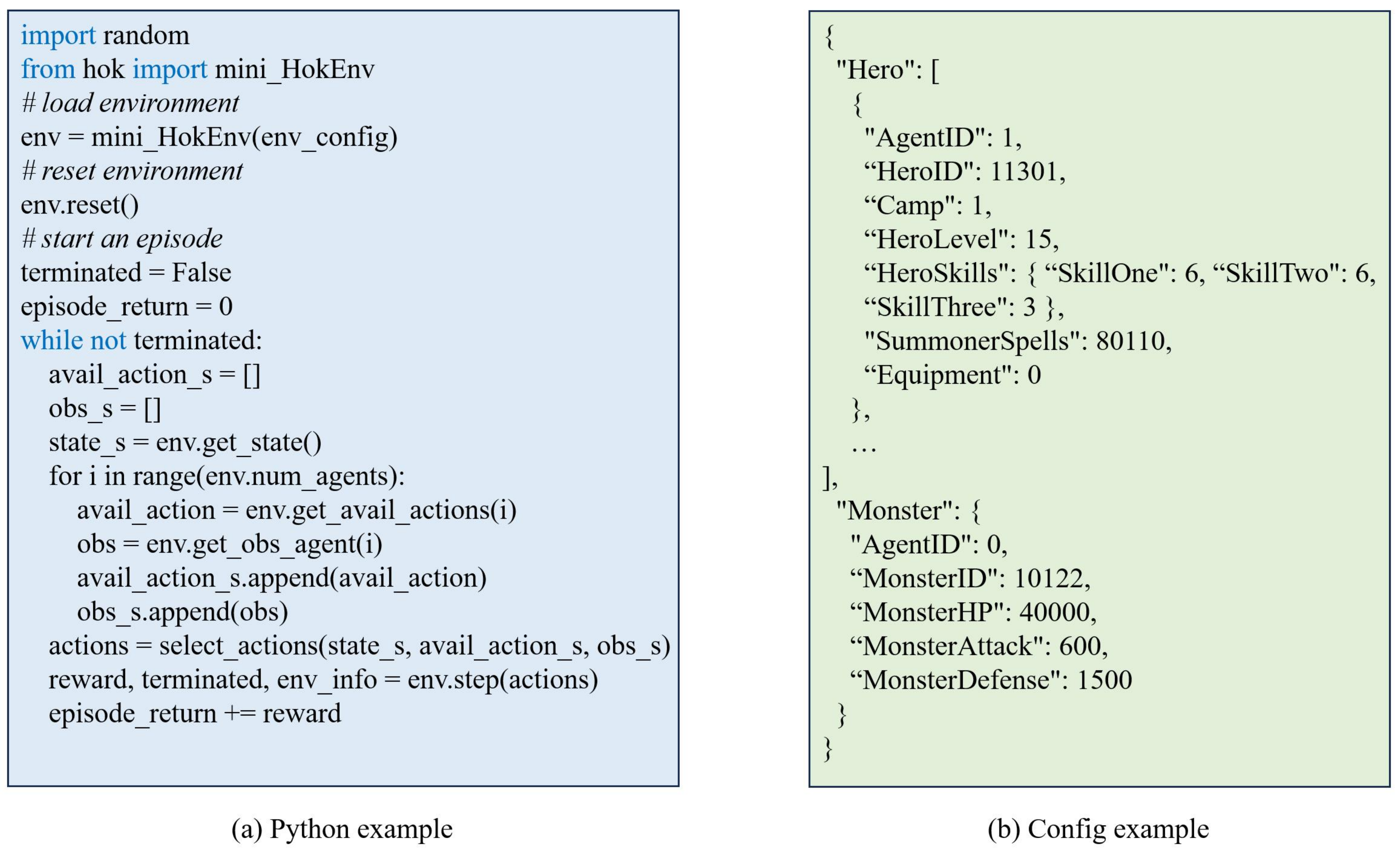}
	\caption{Python example and config example.
    }
	\label{config}
\end{figure*}

\begin{itemize}
\item \textbf{reset():} This function starts a new episode, initializing and storing information for the game.

\item \textbf{get\_state(); get\_obs(); get\_avail\_actions():} These functions retrieve joint observations, individual observations, and available actions for the multi-agent system, as elucidated in Section~\ref{environment}. 
Notably, the observations and available actions are vectorized to encompass information for all agents.

\item \textbf{select\_actions():} This function obtains the decision of each agent, typically derived from models trained by MARL algorithms.

\item \textbf{step():} This function is used to transmit each agent's decision to the simulator and update the state information of environment.
It also returns the reward, episode termination flags, and debugging information. 
Additionally, it encompasses a subordinate function to convert actions into executable commands for the environment.
\end{itemize}
To be noted, the encapsulation provides a clear and accessible interface conducive to the swift development of novel algorithms.
Users retain considerable flexibility to tailor these functions according to their specific requirements.

Moreover, the underlying engine of our benchmark environment is crafted with highly optimized code, allowing it to operate efficiently even on standard laptop configurations.
This efficiency surpasses that of comparable MARL benchmarks, which often require server-level resources to run. Consequently, our environment is notably lightweight and accessible.
Leveraging a solitary CPU core, the environment executes a full episode within a mere second, thereby achieving a performance of approximately 0.5 million samples per hour.
Such expedited processing capabilities afford researchers with limited computational resources the ability to swiftly assess the efficacy of MARL algorithms.

\subsection{Efficiency Analysis}
To further illustrate the efficiency of our Mini HoK environment, we conduct analysis of the training resources required for our environment versus other environments based on the game Honor of Kings.
Given the significant differences in tasks across environments, we focus on the computing resources utilized when running experiments, as detailed in the accompanying Table~\ref{efficiency}.
It can be observed that our Mini HoK operates with substantially higher efficiency compared to its counterparts.
Notably, while Hok Arena and Hok 3v3 necessitate distributed reinforcement learning techniques to enhance training efficiency, our environment does not require such techniques, underscoring its inherent efficiency.

Moreover, our experiments utilized simple and general feature designs, facilitating the use of straightforward network architectures.
In contrast, other environments, due to their complexity, necessitate elaborately designed features with extensive information, resulting in higher-dimensional input vectors and more intricate network architectures.
In summary, the efficiency of our Mini HoK environment is exceptional. 
Despite its simplicity and lightweight characteristic, the environment still offers ample improvement space for MARL algorithms, inviting further research and exploration.

\begin{table}[htbp]
\centering
\resizebox{\textwidth}{!}{
\renewcommand{\arraystretch}{1.3}
{
\begin{tabular}{|c|c|c|c|c|c|c|}
\hline
          & Observation Space & Action Space & CPU Cores     & GPU & Training Hours & Final Result \\ \hline
Hok Arena \cite{wei2022honor} & 491               & 83 & 128 & NVIDIA Tesla V100 &  6.16       & Beat BT AI             \\ \hline
Hok 3v3 \cite{liu2023hokv}   & 4586              & 161 & 64  & NVIDIA Tesla T4 &54.30      & Beat BT AI              \\ \hline
Mini HoK  & 6                 & 13 & 4   & Nvidia GeForce RTX 4090 & 13.21        & Converge               \\ \hline
Mini HoK  & 6                 & 13 & 40   & / & 15.41          & Converge                 \\ \hline
\end{tabular}
}}
\caption{Efficiency analysis of environments based on game Honor of Kings, including the training resources required to run the environments and the observation and action spaces, which influence the complexity of neural networks adopted.
We also include the training outcomes in these environments, where Hok Arena and Hok 3v3 reports the training duration until the system defeats behaviour-tree (BT) opponents while our experiments focus on the time required to achieve convergence.}
\label{efficiency}
\end{table}

\subsection{Experiment Results}
As a reference, we evaluate the performance of VDN \cite{sunehag2017value}, QMIX \cite{rashid2018qmix}, QATTEN \cite{yang2020qatten}, QPLEX \cite{wang2020qplex}, MAPPO \cite{yu2022surprising} and HAPPO \cite{kuba2021trust} in the environment, all of which are introduced in supplement materials.
In order to facilitate these evaluations, we provide example code based on the Pymarl2 algorithm library \cite{hu2021rethinking}.
Given the significant variability among heroes in the Honor of Kings game, the basic hero configuration we adopt consists of five heroes, namely Zhuang Zhou, Di Renjie, Diao Chan, Sun Wukong and Cao Cao, comprising a standard five-hero team of the game.
Performance is evaluated based on the system's ability to engage in combat, measured by the total number of damage the system can cause. 
This metric effectively circumvents the limitation in measuring the upper bound of algorithm performance, as the maximum health points of the Dragon can be adjusted.

As our benchmark is a lightweight environment, the multi-agent system can swiftly acquire relevant knowledge and we run every algorithm for 10000 episodes across 5 random seeds.
The observation feature adopted in the algorithms follows the descriptions in Section~\ref{environment}.
To ensure fair comparisons among different algorithms, each agent's network utilized the same architecture. 
Further technical details of the experiments can be found in the appendix.

In order to assess the system's performance under varying configurations, we devise several experimental setups.
To be noted, although quite a few parameters of the environment can be adjusted, our benchmark is built on the Honor of Kings game engine, necessitating adherence to certain foundational principles.
For instance, in the game heroes can upgrade their skills once every time they level up, which can result in increased damage and reduce the cooldown time.
Skill 1 and skill 2 can undergo six upgrades, whereas skill 3 can only be upgraded when the hero's level is a multiple of 4.
In our experiments, we have the flexibility to configure heroes to abstain from learning skills and only utilize normal attacks, even when their levels are set to 15 but it should be noted that skill 3 cannot be acquired until the hero attains level 4. 
In addition to parameter adjustments, our environment allows for the configuration of different hero compositions to examine the impact of heterogeneity in MARL. 
Specifically, we conduct experiments using five identical heroes to investigate the effects of homogeneous team configurations.

Based on these considerations, the settings of the conducted experiments are shown in table~\ref{tab1}, illustrating the levels of heroes, their respective skill levels, hero compositions and whether they possess equipment.
To be noted, each experimental configuration maintains consistent settings for all heroes to ensure simplicity and coherence, namely, the skill levels and whether to have equipment are uniform for each hero within a given setting.
Moreover, under current setting, a hero equipped with equipment possesses six pre-set equipment, which is determined by the game engine.

\begin{figure*}[htbp]
	\centering
	\includegraphics[width=1.0\textwidth]{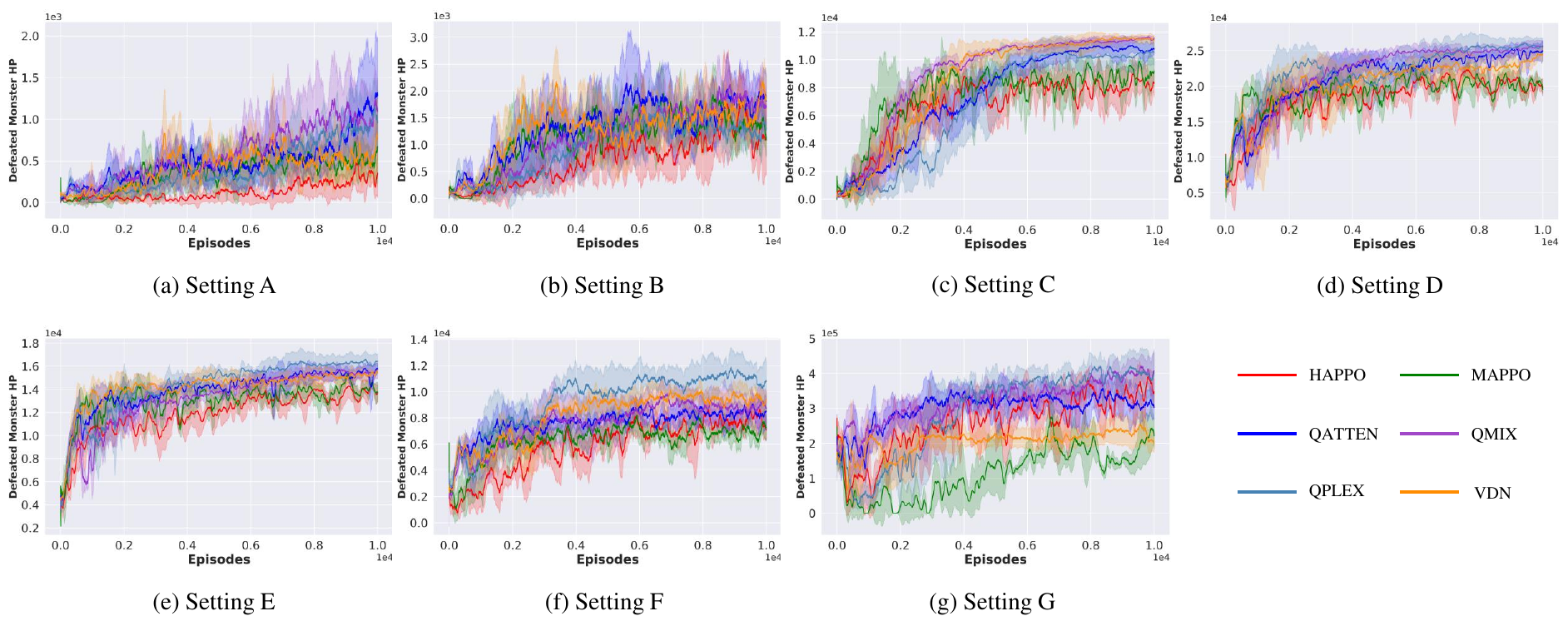}
	\caption{Experiment results for different settings that reflects the damage that the system can cause during the training. 
    The solid lines show the average performance across 5 random seeds and the shaded areas reflect the standard error of the performance outcomes.}
	\label{res}
\end{figure*}




\begin{table*}[htbp]
\centering
\resizebox{\textwidth}{!}{
\renewcommand{\arraystretch}{1.25}
\begin{tabular}{|c|c|c|c|c|c|c|c|}
\hline
\diagbox{Setting}{Method}      & VDN \cite{sunehag2017value}    & QMIX \cite{rashid2018qmix}   & QATTEN \cite{yang2020qatten}   & QPLEX \cite{wang2020qplex} & MAPPO \cite{yu2022surprising} & HAPPO \cite{kuba2021trust} & Rule   \\\hline
A & 808 $\pm$ 391 & 1099 $\pm$ 343 & 1280 $\pm$ 661 & 964 $\pm$ 358 & 
664 $\pm$ 434 & 349 $\pm$ 294 & 3806\\\hline
B & 1830 $\pm$ 473 & 1726 $\pm$ 360 & 1810 $\pm$ 450 & 1309 $\pm$ 208 & 
1115 $\pm$ 315 & 1088 $\pm$ 649 & 5101\\\hline
C  & 11399 $\pm$ 385 & 11562 $\pm$ 59 & 10753 $\pm$ 426 & 10548 $\pm$ 922 & 
9187 $\pm$ 879 & 8344 $\pm$ 978 & 12313\\\hline
D  & 24465 $\pm$ 960 & 25571 $\pm$ 835 & 24973 $\pm$ 1343 & 25661 $\pm$ 1064 & 
19545 $\pm$ 562 & 19906 $\pm$ 1224 & 28133\\\hline
E  & 15614 $\pm$ 123 & 15606 $\pm$ 685 & 15750 $\pm$ 426 & 16427 $\pm$ 670 & 
13769 $\pm$ 858 & 13698 $\pm$ 733 & 18131\\\hline
F  & 9458 $\pm$ 674 & 8709 $\pm$ 1268 & 8502 $\pm$ 1011 & 10814 $\pm$ 1481 & 
7261 $\pm$ 830 & 7126 $\pm$ 825 & 15556\\\hline
G  & 198927 $\pm$ 26453 & 408239 $\pm$ 49792 & 303081 $\pm$ 34428 & 401821 $\pm$ 67922 & 221668 $\pm$ 37440 & 348075 $\pm$ 34815 & 500000\\\hline
\end{tabular}}
    \caption{The caused damage of the multi-agent system under different settings using different methods.
    To be noted, we also provide a method based on heuristic rules for better comparison.
  We report the average performance across 5 random seeds and the standard errors.}
    \label{final_res}
\end{table*}

\begin{figure*}[htbp]
	\centering
	\includegraphics[width=0.5\textwidth]{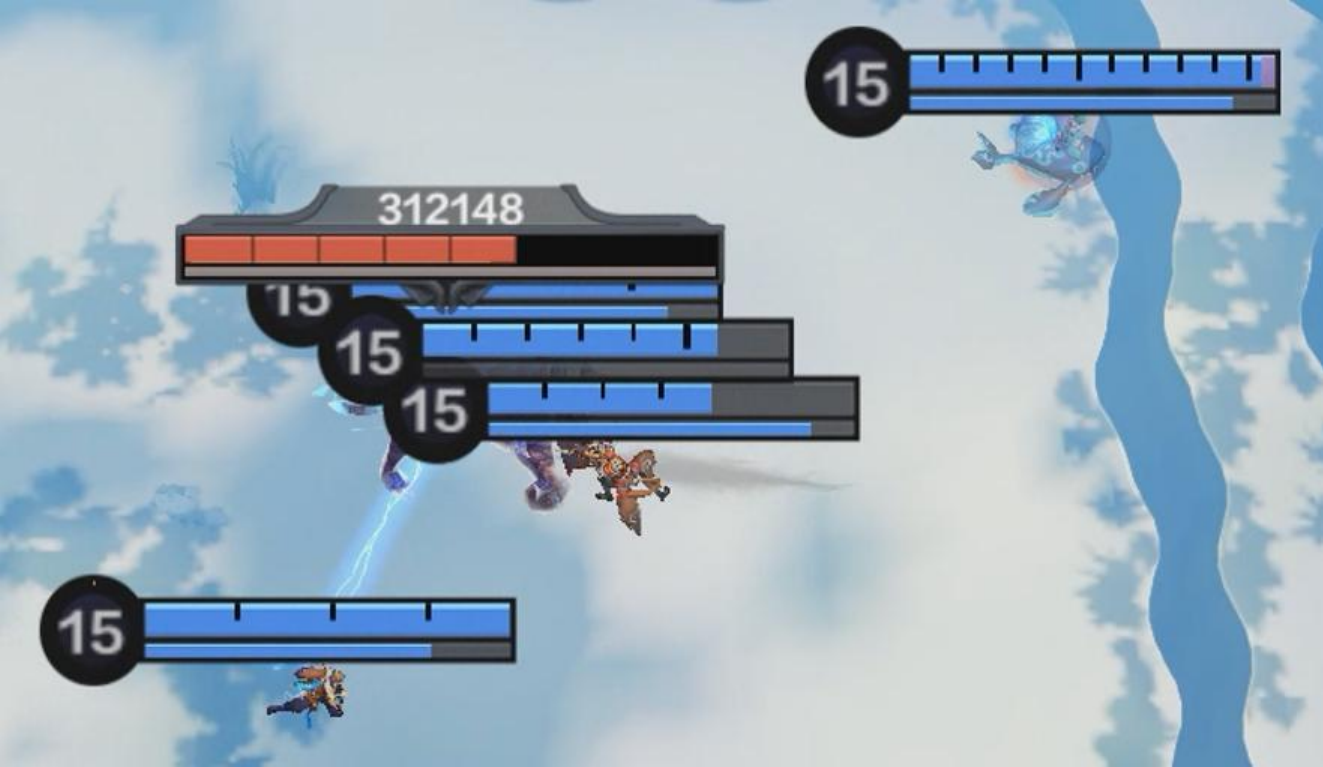}
	\caption{The lazy agent scenario.
    }
	\label{lazy_agent}
\end{figure*}

\subsection{Experiment Analysis}
The experimental results for each setting are presented in figure~\ref{res} and table~\ref{final_res}. 
In addition to the classic MARL algorithms, we include a method based on heuristic rules for better comparison. 
The results indicate that agents with higher hero levels, enhanced skill levels, and equipment inflict more damage. 
Furthermore, the basic hero composition, which includes a variety of hero types, outperforms homogeneous team configurations, underscoring the potential of heterogeneous agents in MARL problems, as diverse agents can assume different roles and enhance cooperation.
Additionally, value decomposition algorithms tend to outperform policy gradient MARL algorithms in our environment, yet they still fall short of surpassing the heuristic-based method.
Despite employing straightforward and universal feature designs in our experiments, which could potentially constrain the efficacy of MARL algorithms, the results still indicate that current MARL algorithms are not fully capable to master the environment. 
This observation highlights significant opportunities for improvement in future MARL research endeavors.

In this experiment, we observed that Zhuang Zhou consistently maintained a certain distance from the monster and did not participate in the actual attack, while the other four agents were actively engaging the monster, as shown in Figure ~\ref{lazy_agent}. 
By reviewing the battle replay video and analyzing Zhuang Zhou's position data and behavior trajectory, we found that Zhuang Zhou did not inflict any damage on the monster throughout the entire attack, exhibiting clear signs of lazy behavior.

The reason behind Zhuang Zhou's inaction may be attributed to his role as a support character, which inherently limits his ability to cause obvious damage to the monster. 
This observation highlights a critical issue in multi-agent collaborative tasks: the emergence of lazy behavior among agents. 
Addressing this challenge is crucial for advancing MARL research.

\section{Conclusion} \label{conclusion}
In this study, we introduce a map editor for Honor of Kings and develop a lightweight MARL environment, Mini Honor of Kings, utilizing this tool.
Constructed on the foundation of the game Honor of Kings and a highly optimized simulator, our environment operates with exceptional efficiency, making it accessible to researchers with limited computing resources.
We provide example code based on the popular Pymarl2 algorithm library and conduct evaluations using several classic MARL algorithms to validate the utility of our benchmark and enable users to swiftly engage with the environment.
The results reveal ample room for improvement for common MARL algorithms within the environment, awaiting further investigation. 
With the flexible map editor, users can freely customize the environment and design diverse scenarios according to their requirements.
We are enthusiastic about sharing the map editor and environemnt with the broader community and anticipate positive contribution to MARL research endeavors.

\appendix


\setcounter{section}{0}
\renewcommand{\thesection}{\Alph{section}}
\renewcommand{\thesubsection}{\thesection.\arabic{subsection}}
\renewcommand{\thesubsubsection}{\thesubsection.\arabic{subsubsection}}

\section{Environment Details}

\subsection{Setup}

An example of the experimental setup is shown in Table~\ref{setup}, the environment contains five heroes and the Dark Dragon. The default five heroes are Zhuang Zhou, Di Renjie, Diao Chan, Sun Wukong, Cao Cao. They have different attributes, for example, Zhuang Zhou has high health points while Di Renjie has low health points. The normal attack range of each hero is also different. In addition, each hero has three skills, and these skills can be divided into four types: ``obj\textunderscore skill'', ``dir\textunderscore skill'', ``pos\textunderscore skill'' and ``talent\textunderscore skill''.

\begin{table}[htb]
\centering
\resizebox{\textwidth}{!}{
\renewcommand{\arraystretch}{1.3}
{
\begin{tabular}{|c|c|c|c|c|c|c|c|}
\hline
\multicolumn{8}{|c|}{5Heroes VS Dark
Dragon} \\
\hline
 & name & ID & Health Points & Normal Attack Range & Skill1 Type & Skill2 Type & Skill3 Type \\
\hline
\multirow{5}{*}{Heroes} & ZhuangZhou & 11301 & 14572 & 2800 & dir\textunderscore skill & obj\textunderscore skill & obj\textunderscore skill \\
\cline{2-8}
& Di Renjie & 13301 & 5706 & 8000 & dir\textunderscore skill & dir\textunderscore skill & dir\textunderscore skill \\
\cline{2-8}
& Diao Chan & 14101 & 8409 & 6000 & dir\textunderscore skill & dir\textunderscore skill & obj\textunderscore skill \\
\cline{2-8}
& Sun Wukong & 16701 & 8743 & 3000 & obj\textunderscore skill & dir\textunderscore skill & obj\textunderscore skill \\
\cline{2-8}
& Cao Cao & 12801 & 9885 & 2800 & dir\textunderscore skill & dir\textunderscore skill & obj\textunderscore skill \\
\hline
Dark Dragon & - & 12202 & 40000 &  &  &  &  \\
\hline
\end{tabular}
}}
    \vspace{0.2cm}
    \caption{Detailed attributes of the 5 heroes and Dark Dragon in the environment, including name, ID, health points, normal attack range, skill type.}
    \label{setup}
\end{table}

\subsection{Attribute parameters of the environment}
Here are some basic attributes of the environment. In the future, multiple maps may be involved, so it is necessary to specify the ``map\_name''. The ``n\_agents'' parameter represents the number of heroes, allowing for the selection of different numbers of heroes subsequently. Additionally, it is necessary to specify the number of steps per episode during training, which is currently set to 150. There are also parameters such as the action space and observation space dimensions for each agent, as well as attributes like the health points of the Dragon at the end of each episode and the types of skills for the heroes.

\hspace*{2em}\textit{\# Name of the map}\\
\hspace*{2em}self.map\textunderscore name = map\textunderscore name\\
\hspace*{2em}\textit{\# Number of agents}\\
\hspace*{2em}self.n\textunderscore agents = num\textunderscore agents 

\hspace*{2em}\textit{\# The maximum number of steps in an episode}\\
\hspace*{2em}self.episode\textunderscore limit = episode\textunderscore limit\\
\hspace*{2em}\textit{\# Step count for each episode}\\
\hspace*{2em}self.\textunderscore episode\textunderscore steps = 0

\hspace*{2em}\textit{\# The action space dimension is 13}\\
\hspace*{2em}self.action\textunderscore space = 13\\
\hspace*{2em}\textit{\# The number of executable actions}\\
\hspace*{2em}self.n\textunderscore actions = self.action\textunderscore space

\hspace*{2em}\textit{\# The observation space dimension is 6}\\
\hspace*{2em}self.observation\textunderscore space = 6\\
\hspace*{2em}\textit{\# Initialize state and observation variables}\\

\hspace*{2em}\textit{\# Save the health points of the Dragon in the current episode}\\
\hspace*{2em}self.monster\textunderscore hp\textunderscore total = []\\
\hspace*{2em}\textit{\# The offset after the movement is discretized is 1500 based on the original coordinates.}\\
\hspace*{2em}self.pos\textunderscore delta = 1500\\
\hspace*{2em}\textit{\# Four skill types}\\
\hspace*{2em}self.SKILL\textunderscore TYPE = ["obj\textunderscore skill", "dir\textunderscore skill", "pos\textunderscore skill", "talent\textunderscore skill"]\\

\subsection{Map}

\begin{figure*}[htb]
	\centering
	\includegraphics[width=0.6\textwidth]{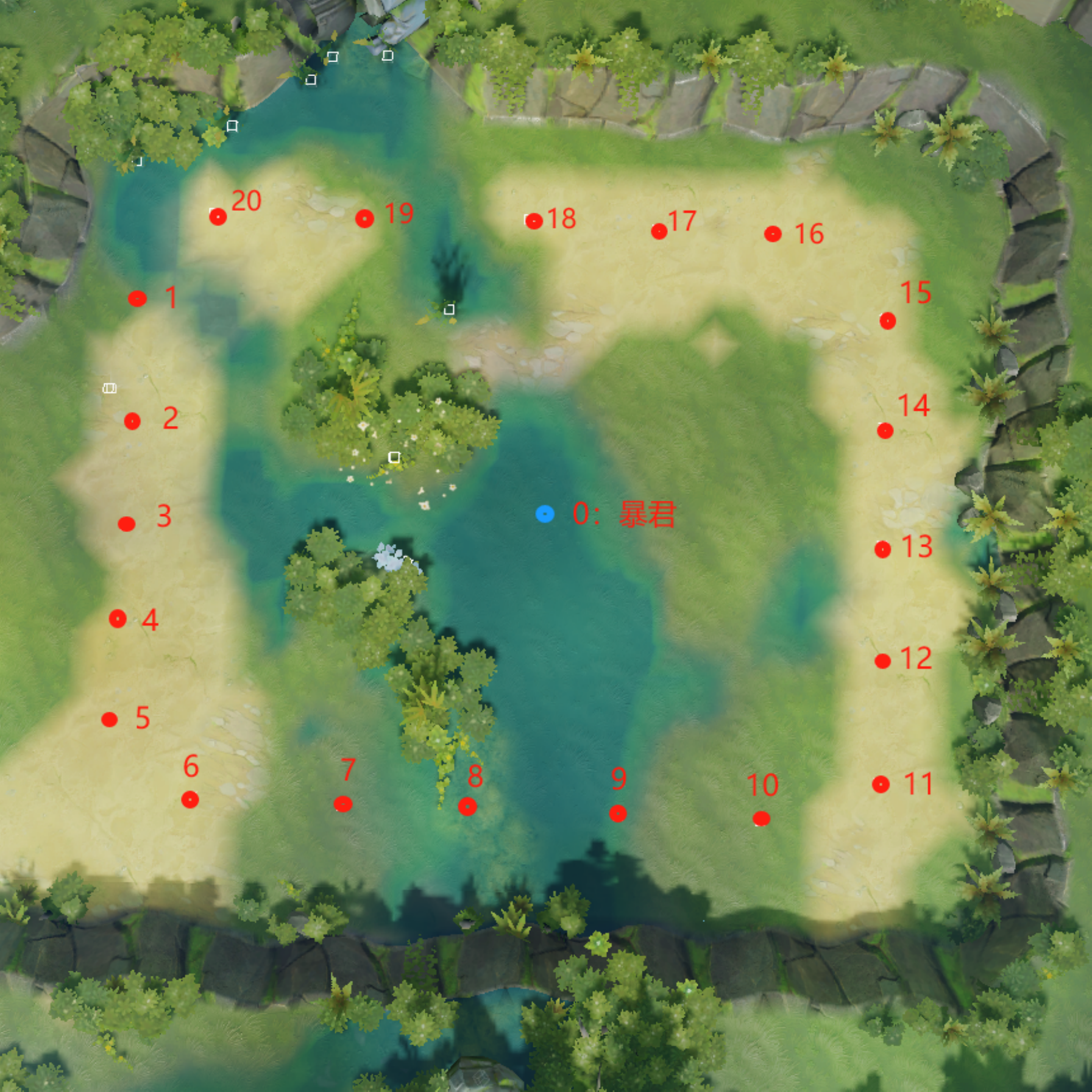}
	\caption{Aerial view of the map.
    }
	\label{map}
\end{figure*}

As illustrated in the figure~\ref{map}, this is an overhead view of the map. At the start of the game, the Dragon spawns in the central position of the map. Surrounding the Dragon are 20 distinct points, and each hero can choose to spawn at any one of these points.

\subsection{Reset}

\begin{figure*}[htb]
	\centering
	\includegraphics[width=0.8\textwidth]{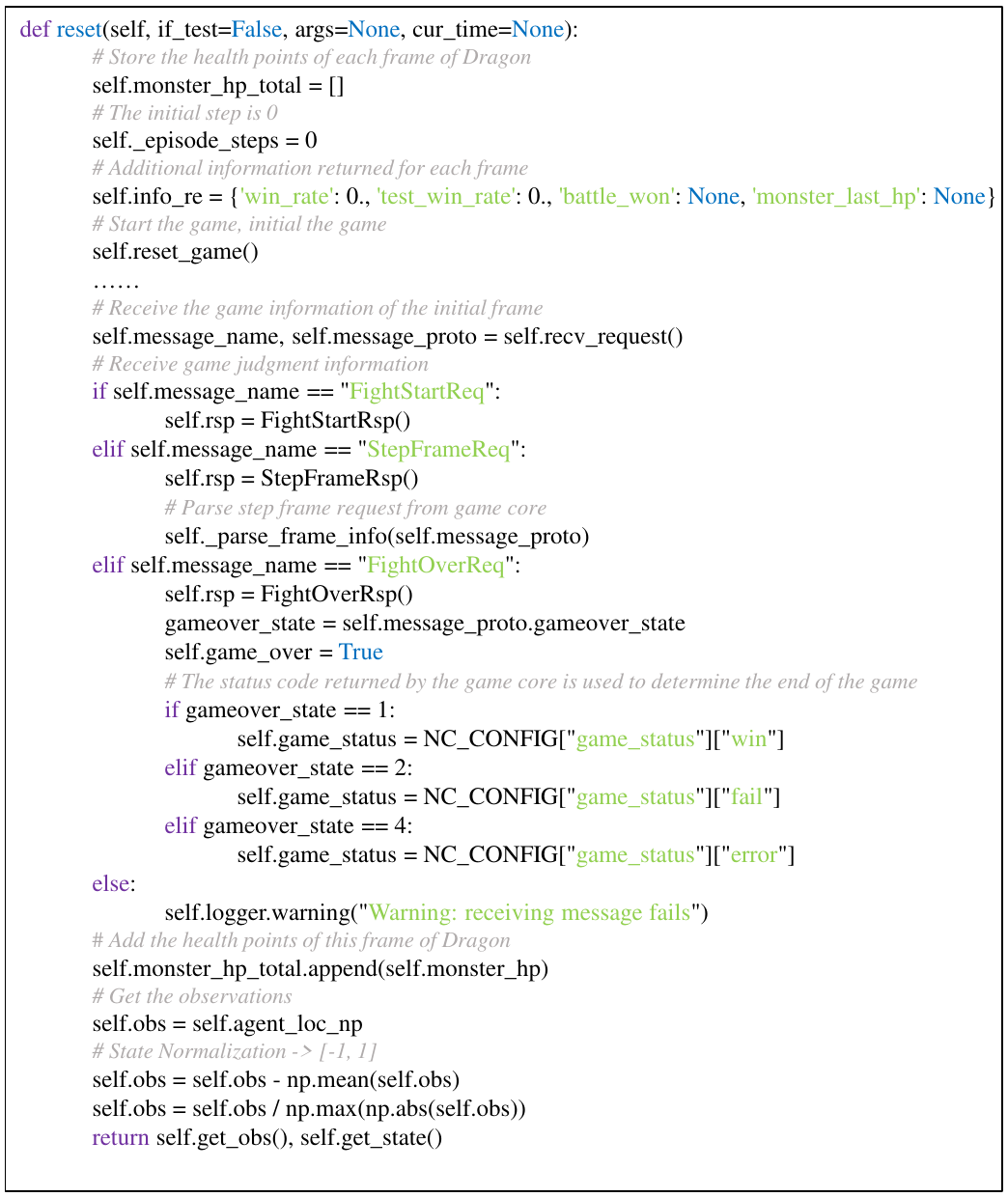}
	\caption{Python example of environment initialization.
    }
	\label{reset}
\end{figure*}

Figure~\ref{reset} presents a python example of initializing the environment. Before the game starts, it initializes various parameters such as health points of the Dragon and the number of game steps. Once the game begins, the client receives game data from the server, which provides the observation and state information for each agent.

\subsection{State}

\begin{figure*}[htbp]
	\centering
	\includegraphics[width=0.5\textwidth]{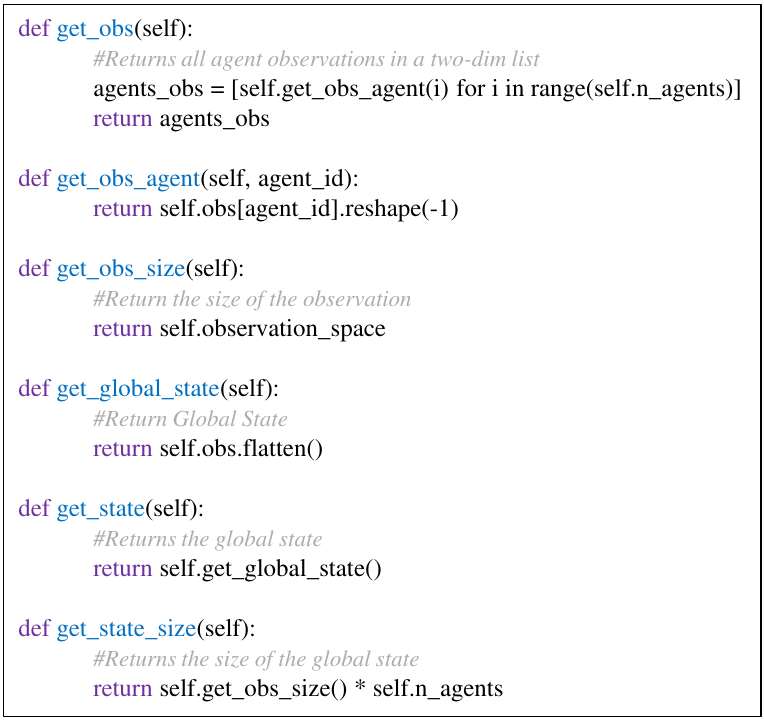}
	\caption{Python example for obtaining observations and states.
    }
	\label{getstate}
\end{figure*}

Figure~\ref{getstate} illustrates a python example for obtaining observations and states. During each step of an episode, the following function is called to retrieve the observation and state information for each agent. This information is then used for decision-making for the next step.

\begin{figure*}[htbp]
	\centering
	\includegraphics[width=0.6\textwidth]{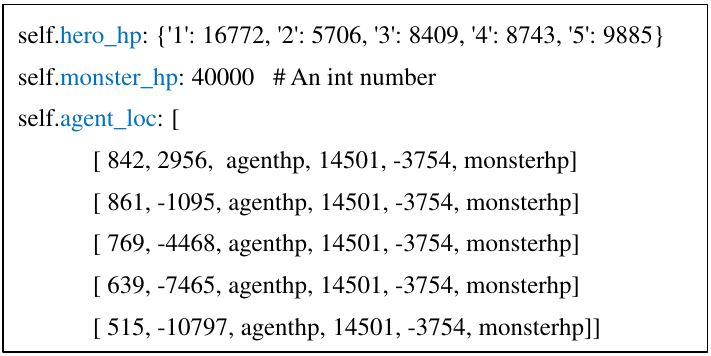}
	\caption{Information used in training.
    }
	\label{usestate}
\end{figure*}

As shown in Figure~\ref{usestate}, during actual training, the observation space dimension of each agent is 6, which is a one-dimensional vector of length 6. The specific information includes six pieces of information: its own x coordinate, its own z coordinate, its own health points, x coordinate of the Dragon, z coordinate of the dragon, Dragon's health points. 
The state is to concatenate the observations of each agent. There are 5 agents in total, and the observation space of each agent is 6, so state is a one-dimensional vector with a length of 30.

\begin{figure*}[htb]
	\centering
	\includegraphics[width=0.95\textwidth]{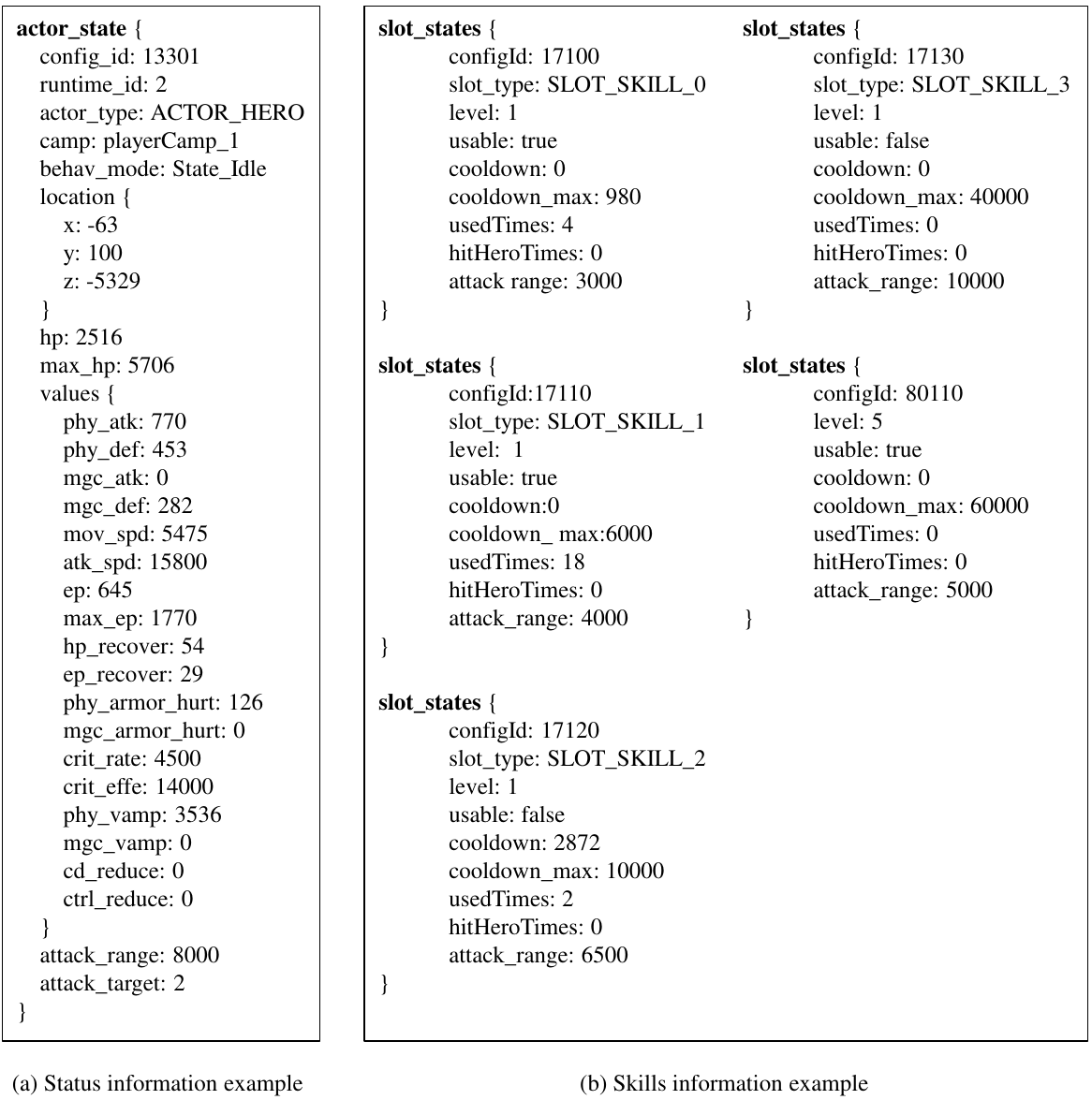}
	\caption{Example of all the information a hero can obtain from the environment.
    }
	\label{allstate}
\end{figure*}

In fact, the server returns a wealth of information. Figure~\ref{allstate} provides an example of all the information a hero can obtain at a specific step, mainly including two parts: hero status and skill status. Besides the agent's coordinates and its health, the information includes physical attack power, magic attack power, health recovery, cooldown reduction, skill levels, and more. This comprehensive data provides robust support for the future expansion of the environment.

\subsection{Actions}
The action space dimension of each agent is 13, which is a one-dimensional vector with a length of 13. The specific action information includes: up, down, left, right, normal attack, skill1, skill2, skill3, skill4, left\textunderscore up, right\textunderscore up, right\textunderscore down, left\textunderscore down. The first 4 actions and the last 4 actions are the movement actions of the agent, which are moving up, moving down, moving left, moving right, moving to the upper left, moving to the upper right, moving to the lower right, moving to the upper left. The 5th to 9th actions are attack actions, respectively normal attack, skill1, skill2, skill3, and summoner skill. 
Normal attack can only be performed within the attack range. Both hero skills and summoner skill have cooldown time, during which the skills are unavailable. The cooldown time of skills is related to the level and equipment of the hero. The higher the level of the hero, the shorter the cooldown time of the skills. Some equipment can also reduce the cooldown time of skills.

\begin{figure}[htb]
	\begin{minipage}{0.48\linewidth}
		\vspace{3pt}
		\centerline{\includegraphics[width=\textwidth]{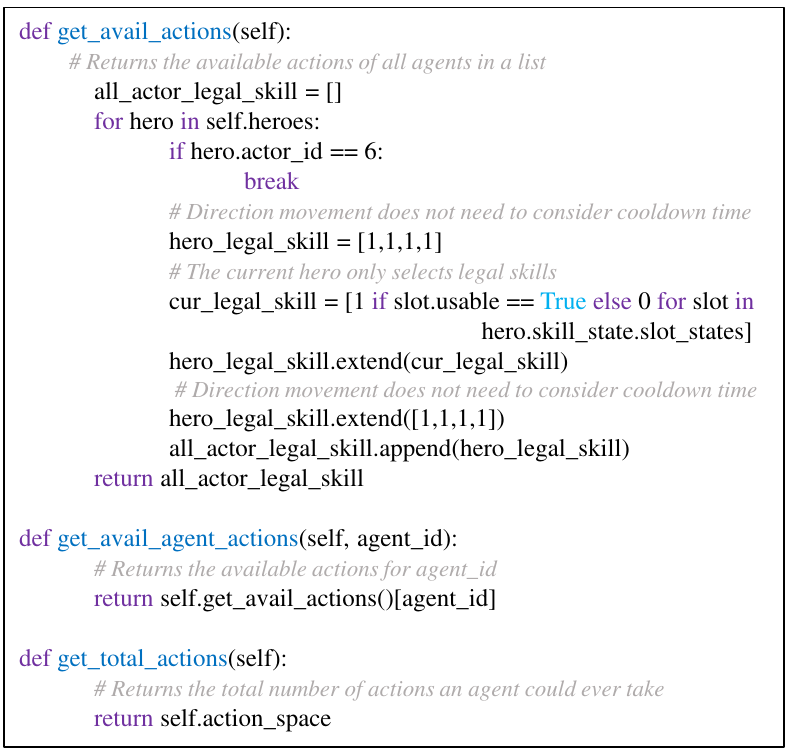}}
		\centerline{Obtaining available actions.}
        \label{getaction}
	\end{minipage}
	\begin{minipage}{0.5\linewidth}
		\vspace{3pt}
		\centerline{\includegraphics[width=\textwidth]{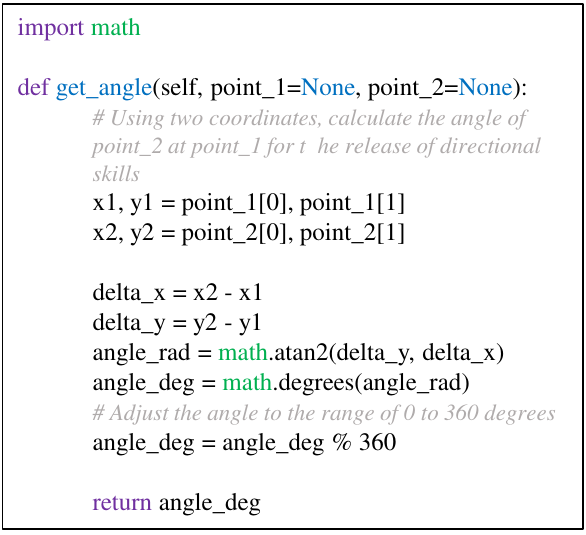}}
		\centerline{Environment interaction.}
	\end{minipage}
	\label{getangle}
 \caption{Python examples for different scenarios. }
 \label{examples}
\end{figure}

As shown in Figure~\ref{examples}, before executing an action, each agent needs to query the available actions at this moment and filter out the unavailable actions. A list with a length equal to the action space, which is 13, is then returned. Each index position represents whether the action can be executed, and 0/1 represents cannot/can be executed respectively.

As shown in Figure~\ref{examples}, before executing directional skills, it is necessary to calculate the direction angle of the Dragon relative to the agent, and then release the skill in the direction of the Dragon.

\subsection{Step}
\begin{figure}[htb]
	\begin{minipage}{0.55\linewidth}
		\vspace{3pt}
		\centerline{\includegraphics[width=\textwidth]{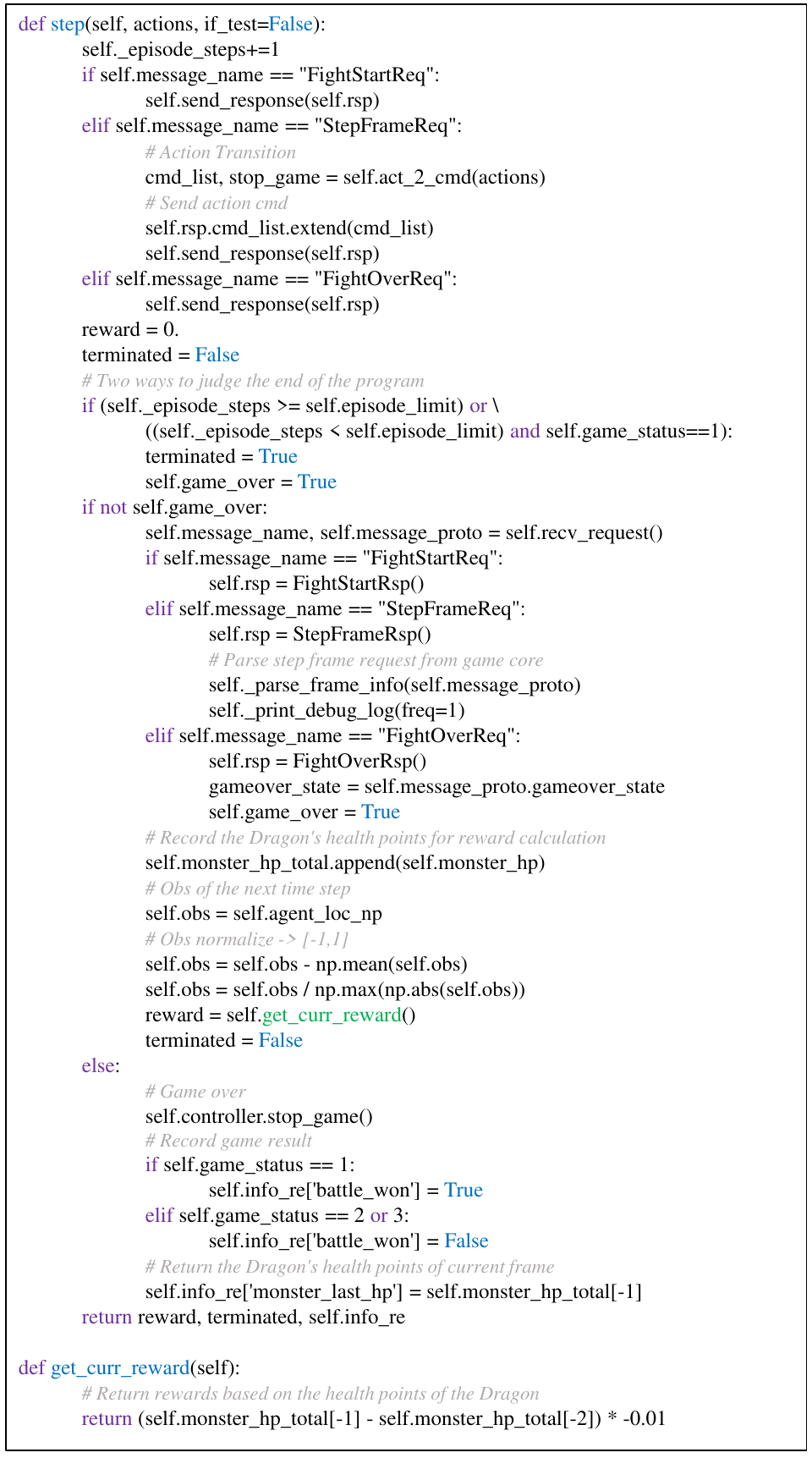}}
		\centerline{Environment interaction.}
	\end{minipage}
	\begin{minipage}{0.4\linewidth}
		\vspace{3pt}
		\centerline{\includegraphics[width=\textwidth]{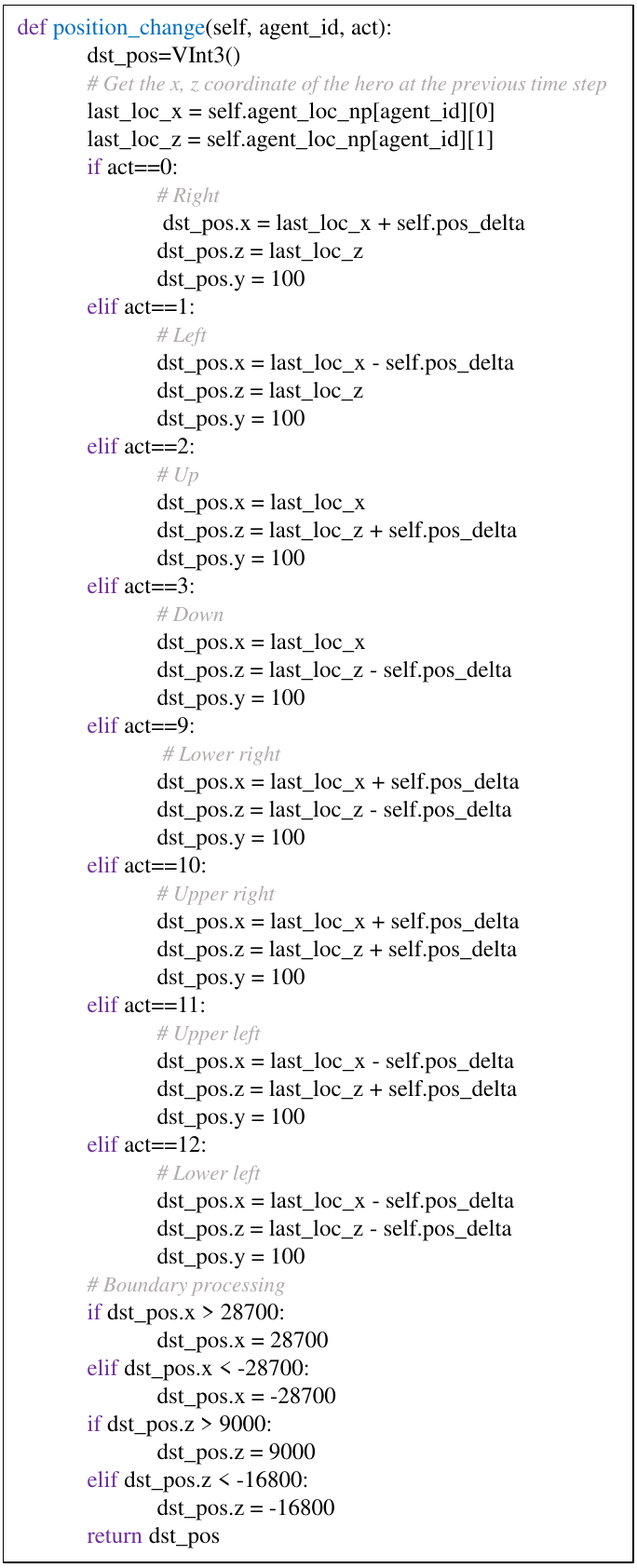}}
		\centerline{Coordinates Calculation after moving.}
	\end{minipage}
 \caption{Python examples for environment interaction and calculating the coordinates after moving. }
 \label{examples2}
\end{figure}

As illustrated in the figure~\ref{examples2}, each interaction with the environment takes the actions of all agents as input and then determines whether the game is over, and parses the information returned by the server. Ultimately, it returns the overall reward, an indicator of whether the game is over, and other state information. It is important to note that the reward is calculated based on the Dragon's health points. The more health the dragon loses, the higher the reward is.

\subsection{Coordinate}
As shown in Figure~\ref{examples2}, this is a python example for calculating the coordinates of an agent after it moves. Based on the action of the agent in the current time step, an offset is obtained, and the coordinates of each agent in the previous time step are added to the offset to obtain the coordinates after the action is performed. There are 8 movement actions, namely up, down, left, right, lower right, upper right, lower left, and upper left.

\begin{figure}[h]
	\centering
	\includegraphics[width=0.8\textwidth]{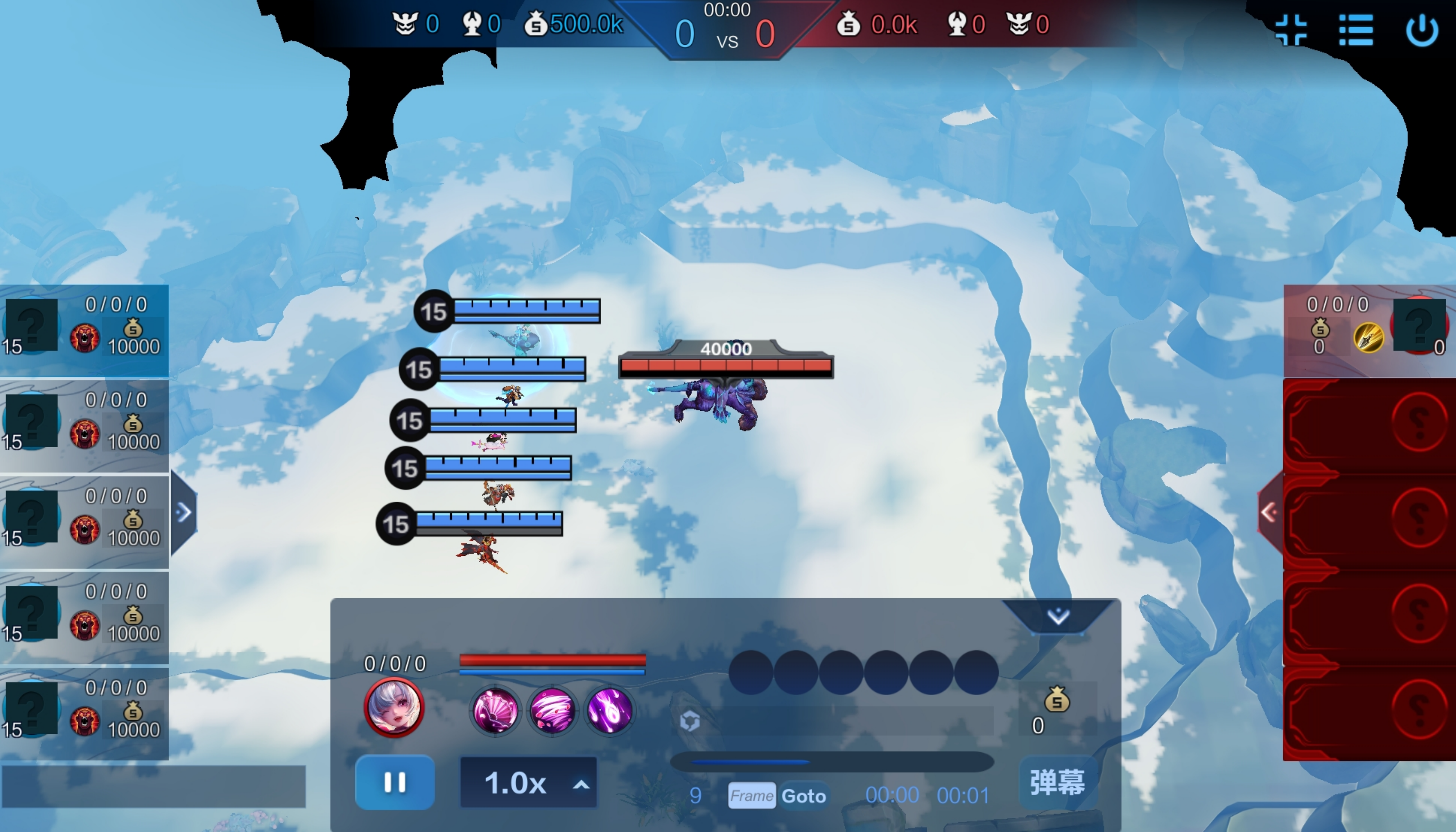}
	\caption{Example of replay screen.}
	\label{replaypic}
\end{figure}

\begin{figure}[htb]
    \centering
    \begin{subfigure}[t]{0.4\textwidth}
        \centering
        \includegraphics[width=\textwidth]{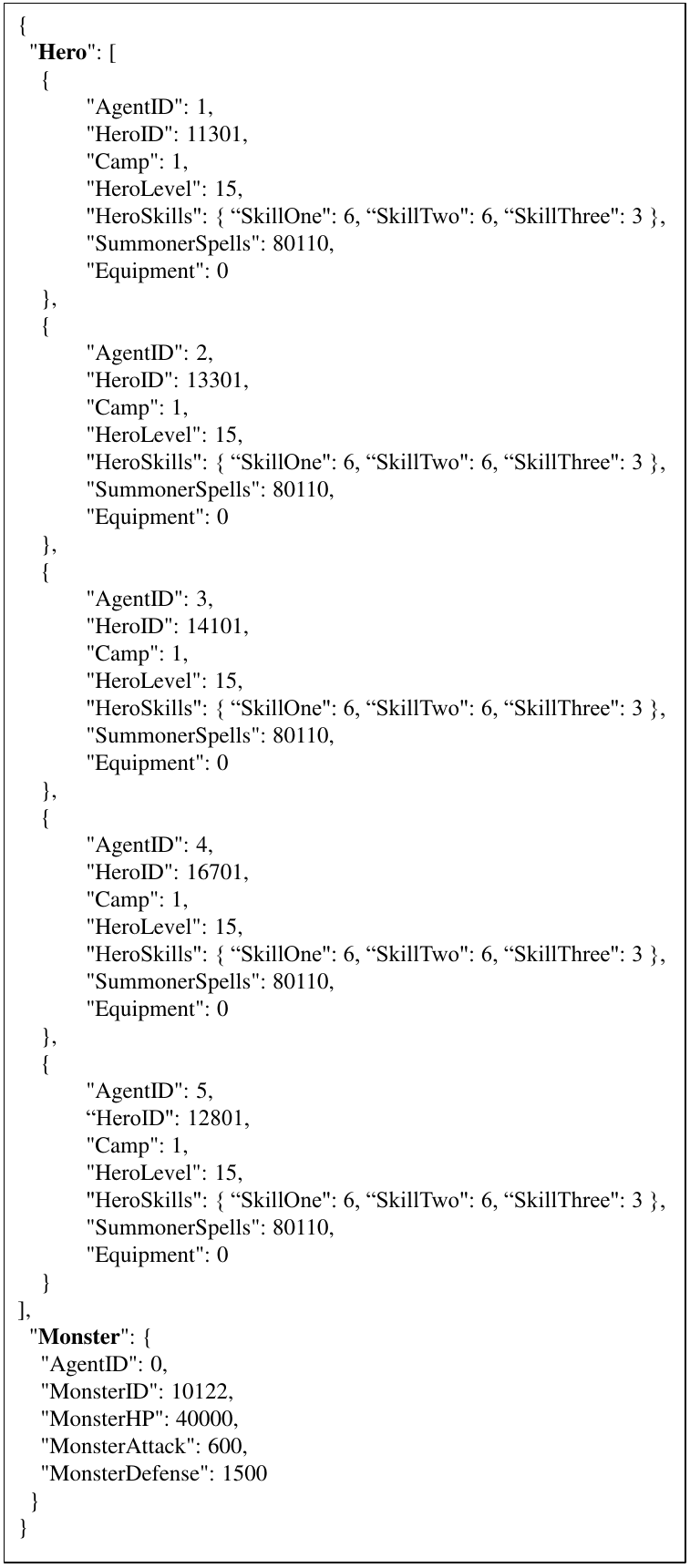}
        \caption{Example of server configuration.}
        \label{serverconfig}
    \end{subfigure}
    \hspace{0.05\textwidth} 
    \begin{subfigure}[t]{0.4\textwidth}
        \centering
        \includegraphics[width=\textwidth]{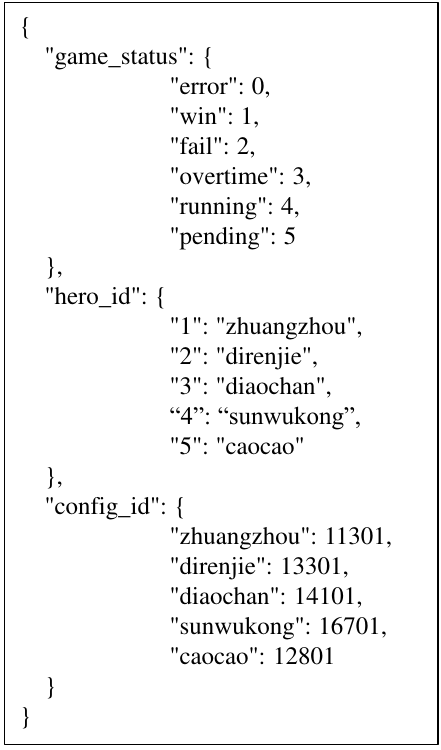}
        \caption{Example of client configuration.}
        \label{clientconfig}
    \end{subfigure}
    \caption{Examples of server and client configurations.}
\end{figure}

\subsection{Replay}

After each episode, the server generates a replay file with an ``abs'' suffix. And there is a Windows-based player that can play these replay files. To play them, simply download the replay files and place them in the appropriate folder on the player. Figure~\ref{replaypic} shows a screenshot of a replay, where you can see the positions, actions, health points, hero levels, skill levels of all heroes, as well as the position, actions, and health points of the Dragon. This clearly illustrates the dynamics between both sides during the match, making it easier and more intuitive for researchers to review the agents' performance and other game information. 

It is important to note that the server's configuration files must match the corresponding configuration files in the player to ensure that the abs file plays normally.

\section{Environment Modes}
As shown in Table~\ref{modes}, this experiment includes seven different modes, each varying in hero levels, skill levels, equipment, and different health points of the Dragon. Compared to other modes, Mode (E) features five identical heroes, which reflects the situation of both homogeneous and heterogeneous. This setup highlights the diversity of the environment and underscores its potential for future expansion.

\begin{table}[htb]
\centering
\renewcommand{\arraystretch}{1.3}{
    \begin{tabular}{|c|c|c|c|c|c|c|}
    \hline
Setting     &   Hero level & Skill 1 & Skill 2  & Skill 3 & Equipment &  \multicolumn{1}{l|}{Hero composition} \\ \hline
(A)   & 1 & 0   & 0  & 0 & \XSolid     & basic    \\  \hline
(B)   & 4 & 0   & 0  & 0 &  \XSolid    & basic    \\ \hline
(C)   & 15 & 0  & 0  & 0 &  \XSolid    & basic    \\ \hline
(D)   & 15 & 6  & 6  & 3 &  \XSolid    & basic    \\ \hline
(E)   & 15 & 6  & 6  & 3 &   \XSolid   & homo (Sun Wukong) \\ \hline
(F)   & 4 & 2   & 1  & 1 &  \XSolid    & basic    \\ \hline
(G)   & 4 & 2   & 1  & 1 &  \Checkmark  & basic      \\ \hline
\end{tabular}}
    \vspace{0.2cm}
    \caption{Different modes of the experiments. Table entries denote the levels of heroes and their respective three skills. 
    The symbol \XSolid~indicates absence of equipment for all heroes, while \Checkmark~signifies all heroes possess six pre-set equipment, as determined by the game engine. 
    When all skill levels are set to zero, it implies that the heroes are only capable of executing normal attacks.
    The basic hero configuration includes the five aforementioned heroes, while the homogeneous hero composition consists of five Sun Wukong for simplicity.}
    \label{modes}
\end{table}

\subsection{Game configuration}
The server configuration allows for the adjustment of various parameters, including hero levels, skill levels, summoner skills, and whether equipment is worn. As shown in Figure~\ref{serverconfig}, the level of each skill of each hero can be set. For the Dragon, attributes such as health points, attack power, and defense can be modified to create different levels of task difficulty. By adjusting server configuration, a wide range of experimental modes can be created to validate various multi-agent tasks. 

Figure~\ref{clientconfig} shows an example of a client configuration, which primarily includes hero ID information that must match the corresponding hero ID information on the server. After updating the hero of the server configuration, you need to make the same changes to the hero in the client configuration. Refer to the hero order in the server configuration.

\subsection{Parameter configuration of 5 heroes and Dragon}
When a hero's level changes, their skills also change accordingly, affecting attributes such as cooldown time, attack power, and recovery. The table~\ref{levels} below list various attribute values for heroes and the Dragon at different levels. It is evident that higher hero levels result in significantly increased values for attributes like health points and attack power.

\begin{table}[htb]
\centering
\resizebox{\textwidth}{!}{
\renewcommand{\arraystretch}{1.3}
{
\begin{tabular}{|c|c|c|c|c|c|c|}
\hline
 Name & Di Renjie & Diao Chan & Sun Wukong & Zhuang Zhou & Cao Cao & Dragon \\
\hline
ID & 13301 & 14101 & 16701 & 11301 & 12801 & 12202 \\
\hline
Type & Hero & Hero & Hero & Hero & Hero & Monster \\
\hline
Camp & 1 & 1 & 1 & 1 & 1 & 2 \\
\hline
Hero Level & 15 / 3 & 15 / 3 & 15 / 3 & 15 / 3 & 15 / 3 &  \\
\hline
Maximum health points & 5706 / 3594 & 5609 / 3389 & 7843 / 5023& 7738 / 3802 & 8185/4753 & 10000 \\
\hline
Physical attack & 449 / 281 & 279 / 183 & 370 / 325 & 296 / 188 & 377 / 221 & 600 \\
\hline
Physical defense& 333 / 129 & 325 /121 & 374 / 134 & 486 / 198 & 412 / 221 & 600 \\
\hline
Magic attack & 0 & 126 & 0 & 0 & 0 & 204 \\
\hline
Magic defense & 162 / 66 & 162 / 66 & 162 / 66 & 162 / 66 & 162 / 66 & 600 \\
\hline
Movement speed & 3607 & 3500 & 3800 & 3900 & 3807 & 3500 \\
\hline
Attack speed bonus & 2800 / 400 & 1400 / 200 & 1400 / 200 & 1400 / 200 & 1423 / 1223 & 0 \\
\hline
Maximum potential energy& 1770 / 570 & 1960 / 700 & 1790 / 620 & 1694 / 602 & 0 & 0 \\
\hline
Health points recover & 54 / 42 & 71 / 47 & 78 / 54 & 94 / 58 & 96 / 60 & 0 \\
\hline
Potential energy recover & 29 / 17 & 31 / 19 & 29 / 17 & 29 / 17 & 0 & 0 \\
\hline
Physical armor hurt & 126 & 0 & 126 & 0 & 126 & 0 \\
\hline
Magic armor hurt & 0 & 42 & 0 & 0 & 0 & 0 \\
\hline
Critical hit rate & 0 & 0 & 2008 & 0 & 0 & 0 \\
\hline
Critical hit effect & 10000 & 10000 & 5042 & 10000 & 10000 & 0 \\
\hline
Physical vamp & 0 & 0 & 0 & 0 & 0 & 0 \\
\hline
Magic vamp & 0 & 0 & 0 & 0 & 0 & 0 \\
\hline
Cooldown time reduction & 0 & 0 & 0 & 0 & 0 & 0 \\
\hline
Control reduce & 0 & 0 & 0 & 0 & 0 & 0 \\
\hline
Attack range & 8000 & 6000 & 3000 & 2800 & 2800 & 3000 \\
\hline
\end{tabular}
}}
    \vspace{0.2cm}
    \caption{Detailed parameter configuration table of heroes and Dragon when the heroes are at level 15 and level 3. To be mentioned, if one attribute of a hero changes when leveling up, we report the respective value at level 15 and 3.}
    \label{levels}
\end{table}

As shown in Table~\ref{summonerskill}, there are 8 summoner skills in total, each with its own specific application scenarios. For the sake of simplicity in our experiments, we have standardized all heroes' summoner skills to ``Berserk''. However, in future studies, it will be possible to assign suitable summoner skills to each hero based on their individual characteristics, which will increase the diversity of the environment.

\begin{table}[htb]
\centering
\resizebox{\textwidth}{!}{
\renewcommand{\arraystretch}{1.3}
{
\begin{tabular}{|c|c|c|p{10cm}|}
\hline
Summoner skill name & Corresponding id & Cooldown time & Introduction \\
\hline
Sprint & 80109 & 90s & Increases movement speed by 30\% for 10 seconds. When activated, removes the slowing effect on the player, reduces the slowing effect by 50\% during sprinting, and increases movement speed by an additional 20\% when out of combat. \\
\hline
Heal & 80102 & 120s & Restore 15\% of your own and nearby teammates' health, and increase nearby friendly forces' movement speed by 15\% for 2 seconds. \\
 \hline
Execute & 80108 & 60s & Immediately causes true damage to nearby enemy heroes equal to 14\% of their lost health. \\
\hline
Berserk & 80110 & 60s & Increases damage by 10\%, toughness by 15\%, physical vamp and magic vamp by 20\% for 7 seconds \\
\hline
Stun & 80103 & 90s & Stuns all nearby enemies for 0.75 seconds and slows them down for 1 second. \\
 \hline
Purify & 80107 & 120s & Removes all negative and control effects on the player and makes them immune to control for 1.5 seconds. \\
\hline
Weaken & 80121 & 90s & Reduces nearby enemy damage by 25\% and increases your own damage reduction rate by 20\% for 4 seconds. \\
\hline
Flash & 80115 & 120s & Move a certain distance in a specified direction. \\
\hline
\end{tabular}
}}
    \vspace{0.2cm}
    \caption{Introduction to summoner skills.}
    \label{summonerskill}
\end{table}

\subsection{Equipment}
As shown in Table~\ref{modes}, for Mode (G), each hero wears equipment that significantly enhances their abilities. For the sake of experimental convenience, we provide each hero with six pieces of equipment, as detailed in Table~\ref{equipment}. In future studies, the types and quantities of equipment each hero carries can be adjusted to create a more diverse environment, thereby accommodating a wider range of task requirements.

\begin{table}[htb]
\centering
\resizebox{\textwidth}{!}{
\renewcommand{\arraystretch}{1.3}
{
\begin{tabular}{|c|c|c|c|c|c|c|}
\hline
Name & Equipment1 & Equipment2 & Equipment3 & Equipment4 & Equipment5 & Equipment6 \\ \hline
ZhuangZhou & Boots of Resistance & Guardian's Glory & Red Lotus Cape & Ominous Premonition & Succubus Cloak & Overlord's Platemail \\ \hline
Di Renjie & Boots of Dexterity & Doomsday & Endless Edge & Sparkforged Dagger & Bloodweeper & Dawn Harvester \\ \hline
Diao Chan & Boots of Tranquility & Book of Blood & Mask of Pain & Glacial Buckler & Eye of the Phoenix & Savant's Wrath \\ \hline
Sun Wukong & Boots of Dexterity & Endless Edge & Master Sword & Muramasa & Claves Sancti & Nightmare \\ \hline
Cao Cao & Boots of Resistance & Spikemail & Nightmare & Claves Sancti & Eye of the Phoenix & Sword of Glory \\ \hline
\end{tabular}
}}
    \vspace{0.2cm}
    \caption{Equipment information for each hero.}
    \label{equipment}
\end{table}

\section{Algorithms and Hyperparameters}
In our experiments, we use 6 different MARL algorithms, including QMIX, VDN, QPLEX, QATTEN, MAPPO and HAPPO. These algorithms use different methods and strategies to optimize the learning process of agents, aiming to solve the decision-making problem of multiple agents in complex dynamic environments.

\subsection{QMIX}
QMIX can be thought of as an extension of DQN to the Dec-POMDP setting. The joint optimal action is found by forcing the joint \( Q \) to adhere to the individual global max (IGM) principle, which states that the joint action can be found by maximising individual agents' \( Q_i \) functions:

\[
\arg \max_{a} Q(s, \tau, a) = \left\{ \begin{array}{l}
\arg \max_{a} Q_1(\tau_1, a_1) \\
\arg \max_{a} Q_2(\tau_2, a_2) \\
\vdots \\
\arg \max_{a} Q_n(\tau_n, a_n)
\end{array} \right.
\]

This central \( Q \) is trained to regress to a target \( r + \gamma \hat{Q}(s, \tau, a) \) where \( \hat{Q} \) is a target network that is updated slowly. The central \( Q \) estimate is computed by a mixing network, whose weights are conditioned on the state, which takes as input the utility function \( Q_i \) of the agents. The weights of the mixing network are restricted to be positive, which enforces the IGM principle by ensuring the central \( Q \) is monotonic in each \( Q_i \). The relevant hyperparameters are shown in Table~\ref{qmix}.
To be mentioned, these are the values in the corresponding configuration file in Pymarl. Mac is the code responsible for marshalling inputs to the neural networks, learner is the code used for learning and runner determines whether experience is collected in serial or parallel.

\subsection{VDN}
VDN (Value Decomposition Networks) decomposes the joint action-value function \( Q(s, \mathbf{a}) \) into a sum of individual value functions:
\[
Q(s, \mathbf{a}) = \sum_{i=1}^n Q_i(s_i, a_i)
\]
This approach allows each agent to independently maximize its own value function, which simplifies the training process. The individual value functions \( Q_i \) are conditioned only on the local observations \( s_i \) and actions \( a_i \) of each agent, promoting decentralization. VDN is particularly useful in scenarios where multiple agents need to work together to achieve a common goal, such as in team-based games or collaborative robotics.The simplicity of VDN makes it easy to implement and extend. The relevant hyperparameters are shown in Table~\ref{vdn}.

\begin{minipage}[c]{0.5\textwidth}
\centering
\setlength{\arrayrulewidth}{1.2pt}
\begin{tabular}{l>{\centering\arraybackslash}m{0.2\linewidth}}
\hline
\setlength{\arrayrulewidth}{0.4pt}
\textbf{Hyperparameters} & \textbf{Value} \\ \hline
Action Selector & epsilon greedy \\
$\epsilon$ Start & 1.0 \\
$\epsilon$ Finish & 0.05 \\
$\epsilon$ Anneal Time & 100000 \\
Runner & episode \\
Batch Size Run & 1 \\
Buffer Size & 5000 \\
Batch Size & 64 \\
Optimizer & Adam \\
$t_{max}$ & 1500000 \\
Target Update Interval & 200 \\
Mac & n\_mac \\
Agent & n\_rnn \\
Agent Output Type & q \\
Learner & nq\_learner \\
Mixer & qmix \\
Mixing Embed Dimension & 32 \\
Hypernet Embed Dimension & 64 \\
Learning Rate & 0.001 \\
td\_lambda  & 0.6 \\
q\_lambda  & False \\ \hline
\label{qmix}
\end{tabular}
\captionof{table}{Hyperparameters used for QMIX experiments. }
\end{minipage}
\begin{minipage}[c]{0.5\textwidth}
\centering
\setlength{\arrayrulewidth}{1.2pt}
\begin{tabular}{l>{\centering\arraybackslash}m{0.2\linewidth}}
\hline
\setlength{\arrayrulewidth}{0.4pt}
\textbf{Hyperparameters} & \textbf{Value} \\ \hline
Action Selector & epsilon greedy \\
$\epsilon$ Start & 1.0 \\
$\epsilon$ Finish & 0.05 \\
$\epsilon$ Anneal Time & 100000 \\
Runner & episode \\
Batch Size Run & 1 \\
Buffer Size & 5000 \\
Batch Size & 64 \\
Optimizer & Adam \\
$t_{max}$ & 1500000 \\
Target Update Interval & 200 \\
Mac & n\_mac \\
Agent & n\_rnn \\
Agent Output Type & q \\
Learner & nq\_learner \\
Mixer & vdn \\
Mixing Embed Dimension & 32 \\
Hypernet Embed Dimension & 64 \\
Learning Rate & 0.001 \\
td\_lambda  & 0.6 \\
q\_lambda  & False \\ \hline
\label{vdn}
\end{tabular}
\captionof{table}{Hyperparameters used for VDN experiments.}
\end{minipage}

\subsection{QPLEX}
QPLEX extends the QMIX framework by using a duplex dueling network architecture. This architecture splits the \( Q \) function into two streams: an advantage stream and a value stream. The advantage stream estimates the relative value of each action, and the value stream estimates the overall value of the state. The final \( Q \) value is computed by combining these two streams:
\[
Q(s, \tau, a) = V(s, \tau) + \left(A(\tau, a) - \frac{1}{|A|} \sum_{a'} A(\tau, a')\right)
\]
By separating the representation of state value and action advantage, QPLEX can more effectively represent complex interactions among agents. The relevant hyperparameters are shown in Table~\ref{qplex}.

\subsection{QATTEN}
QATTEN incorporates attention mechanisms into multi-agent Q-learning. It uses self-attention to compute a weighted representation of each agent's information and cross-attention to capture the interactions between agents. The attention mechanisms allow QATTEN to dynamically adjust the importance of different agents' information, which helps in coordinating joint actions:
\[
Q(s, \tau, a) = \sum_{i=1}^n \alpha_i Q_i(\tau_i, a_i)
\]
where \(\alpha_i\) are attention weights computed based on the agents' states and actions.
The technique can improve coordination among agents, leading to better performance in complex tasks. The relevant hyperparameters are shown in Table~\ref{qatten}.

\begin{minipage}[c]{0.5\textwidth}
\centering
\setlength{\arrayrulewidth}{1.2pt}
\begin{tabular}{l>{\centering\arraybackslash}m{0.2\linewidth}}
\hline
\setlength{\arrayrulewidth}{0.4pt}
\textbf{Hyperparameters} & \textbf{Value} \\ \hline
Action Selector & epsilon greedy \\
$\epsilon$ Start & 1.0 \\
$\epsilon$ Finish & 0.05 \\
$\epsilon$ Anneal Time & 100000 \\
Runner & episode \\
Batch Size Run & 1 \\
Buffer Size & 5000 \\
Batch Size & 64 \\
Optimizer & Adam \\
$t_{max}$ & 1500000 \\
Target Update Interval & 200 \\
Agent Output Type & q \\
Learner & dmaq\_qatten\_learner \\
double\_q & True \\
Mixer & dmaq \\
Mixing Embed Dimension & 32 \\
Hypernet Embed Dimension & 64 \\
adv\_hypernet\_layers & 2 \\
adv\_hypernet\_embed & 64 \\
Learning Rate & 0.001 \\
td\_lambda  & 0.6 \\
num\_kernel & 4 \\
is\_minus\_one & True \\
weighted\_head & True \\
is\_adv\_attention & True \\
is\_stop\_gradient & True \\ \hline
\label{qplex}
\end{tabular}
\captionof{table}{Hyperparameters used for QPLEX experiments. }
\end{minipage}
\begin{minipage}[c]{0.5\textwidth}
\centering
\setlength{\arrayrulewidth}{1.2pt}
\begin{tabular}{l>{\centering\arraybackslash}m{0.2\linewidth}}
\hline
\setlength{\arrayrulewidth}{0.4pt}
\textbf{Hyperparameters} & \textbf{Value} \\ \hline
Action Selector & epsilon greedy \\
$\epsilon$ Start & 1.0 \\
$\epsilon$ Finish & 0.05 \\
$\epsilon$ Anneal Time & 100000 \\
Runner & episode \\
Batch Size Run & 1 \\
Buffer Size & 5000 \\
Batch Size & 64 \\
Optimizer & Adam \\
$t_{max}$ & 1500000 \\
Target Update Interval & 200 \\
Mac & n\_mac \\
Agent & n\_rnn \\
Agent Output Type & q \\
Learner & nq\_learner \\
Mixer & qatten \\
Learning Rate & 0.001 \\
td\_lambda  & 0.6 \\
n\_query\_embedding\_layer1 & 64 \\
n\_query\_embedding\_layer2 & 32 \\
n\_key\_embedding\_layer1 & 32 \\
n\_head\_embedding\_layer1 & 64 \\
n\_head\_embedding\_layer2 & 4 \\
n\_attention\_head & 4 \\
n\_constrant\_value & 32 \\
type & weighted \\
agent\_own\_state\_size & False \\ \hline
\end{tabular}
\label{qatten}
\captionof{table}{Hyperparameters used for QATTEN experiments.}
\end{minipage}

\subsection{MAPPO}
MAPPO (Multi-Agent Proximal Policy Optimization) adapts the single-agent PPO algorithm to the multi-agent setting. It uses a centralized value function to estimate the joint value of all agents' states and actions while maintaining decentralized policies for each agent. MAPPO optimizes a clipped surrogate objective to ensure stable updates:
\[
L^{CLIP}(\theta) = \mathbb{E}_t \left[ \min \left( \frac{\pi_\theta(a_t | s_t)}{\pi_{\theta_{old}}(a_t | s_t)} \hat{A}_t, \text{clip} \left( \frac{\pi_\theta(a_t | s_t)}{\pi_{\theta_{old}}(a_t | s_t)}, 1 - \epsilon, 1 + \epsilon \right) \hat{A}_t \right) \right]
\]
This helps in achieving stable and efficient learning in multi-agent environments. The relevant hyperparameters are shown in Table~\ref{mappo}.

\begin{table}[htb]
\centering
\setlength{\arrayrulewidth}{1.1pt}
\begin{tabular}{l>{\centering\arraybackslash}m{0.2\linewidth}m{0.2\linewidth}}
\hline
\setlength{\arrayrulewidth}{0.4pt}
\textbf{Hyperparameters} & \textbf{MAPPO} & \textbf{HAPPO}\\ \hline
use\_valuenorm & True &True\\
Activation & ReLU & ReLU\\
initialization\_method & orthogonal & orthogonal \\
use\_naive\_recurrent\_policy & False & False \\
data\_chunk\_length & 10  & 10\\
weight\_decay & 0 & 0\\
std\_y\_coef & 0.5 & 0.5\\
value\_loss\_coef & 1 & 1\\
max\_grad\_norm & 10.0& 10.0 \\
GAE lambda & 0.95 & 0.95\\
use\_policy\_active\_masks & True & True\\
action\_aggregation\_& prod & prod\\
accept\_ratio & 0.5 & 0.5\\
use\_proper\_time\_limits & True & True\\
use\_feature\_normalization & True & True\\
Gain & 0.01 & 0.01\\
num\_GRU\_layers & 1 & 1\\
optim\_eps & 1e-5 & 1e-5\\
std\_x\_coef & 1 & 1\\
use\_clipped\_value\_loss  & True & True\\
use\_max\_grad\_norm & True & True\\
Use GAE & True & True\\
use\_huber\_loss & True & True\\
huber\_delta & 10.0 & 10.0\\ 
hidden\_sizes & [128,128] & [128,128]\\
Agent & n\_rnn & n\_rnn\\
Learner & mappo\_learner & happo\_learner\\
num\_env\_steps & 1500000 & 1500000\\
torch\_threads & 1  & 1 \\
ppo\_epoch & 5 & 5\\
clip\_param & 0.2 & 0.2 \\
entropy\_coef & 0.01 & 0.01\\
backtrack\_coef & 0.8 & 0.8\\
use\_linear\_lr\_decay & False & False\\
actor\_lr & 5e-4 & 5e-4\\
critic\_lr & 5e-4 & 5e-4\\
critic\_epoch & 5  & 5\\
actor\_mini\_batch & 1 & 1\\
critic\_mini\_batch & 1 & 1\\
Gamma & 0.99 & 0.99 \\
kl\_threshold & 0.005 & 0.005\\ \hline
\end{tabular}
    \vspace{0.2cm}
    \caption{Hyperparameters used for MAPPO and HAPPO experiments.}
    \label{mappo}
\end{table}

\subsection{HAPPO}
HAPPO (Heterogeneous-Agent Proximal Policy Optimization) extends MAPPO by allowing for heterogeneous agents with different observation spaces, action spaces, and network architectures. HAPPO maintains a centralized critic for the joint value function but uses individual actor networks for each agent. The objective function for HAPPO is similar to MAPPO, ensuring stable updates while allowing for diverse agent capabilities:
\[
L^{HAPPO}(\theta) = \sum_{i=1}^n \mathbb{E}_t \left[ \min \left( \frac{\pi_\theta^i(a_t^i | s_t^i)}{\pi_{\theta_{old}}^i(a_t^i | s_t^i)} \hat{A}_t^i, \text{clip} \left( \frac{\pi_\theta^i(a_t^i | s_t^i)}{\pi_{\theta_{old}}^i(a_t^i | s_t^i)}, 1 - \epsilon, 1 + \epsilon \right) \hat{A}_t^i \right) \right]
\]
HAPPO allows for more flexible and scalable multi-agent learning by accommodating heterogeneity among agents. The relevant hyperparameters are shown in Table~\ref{mappo}.


\section{Characteristics and limitations}

Aiming to contribute to MARL research, Mini Honor of Kings environment offers several notable advantages. Firstly, it is built on a highly optimized simulator, ensuring efficient performance even for researchers with limited computing resources, and is a particularly lightweight environment. Secondly, Honor of Kings is a popular game with many people playing and familiar with it.  Mini Honor of Kings, based on the game mechanism of Honor of Kings and can simulate various complex tasks and challenging scenarios, which helps researchers quickly test and verify the performance of different algorithms in MARL tasks. Furthermore, it includes a flexible map editor that allows researchers to customize the environment and design diverse scenarios, including modifying the hero's level, skills, whether to possess equipment, and health points of the Dragon to meet different research needs. Additionally, in the experimental evaluations, all reinforcement learning algorithms performed worse than the simple rule-based approach, indicating that there is still considerable potential for improvement and optimization in reinforcement learning. This environment, therefore, holds significant research potential.

The main limitation of Mini Hok environment is that only heroes can be controlled by the RL agent. The Dragon is controlled by simple rules with few modifiable attributes, making this environment unsuitable for adversarial scenarios.

\bibliographystyle{unsrt}  
\bibliography{template}  







\end{document}